\documentclass[twocolumn,
showpacs,
preprintnumbers,
amsmath,
amssymb,
groupedaddress,
superscriptaddress
]{revtex4}

\usepackage{graphicx}
\usepackage{dcolumn}
\usepackage{bm}
\usepackage{xcolor}
\usepackage{verbatim}

\usepackage{hyperref}
\usepackage[mathlines]{lineno}

\usepackage{epstopdf}

\begin{document}


\title{Forward-backward multiplicity and momentum correlations \\ in $pp$ collisions at LHC energies}

\author{Mitali Mondal}
\email{mitalimon@gmail.com}
\affiliation{Nuclear and Particle Physics Research Centre, Department of Physics, Jadavpur University, Kolkata - 700032, India}
\affiliation{School of Studies in Environmental Radiation \& Archaeological Sciences, Jadavpur University, Kolkata - 700032, India}

\author{Joyati Mondal}
\affiliation{Nuclear and Particle Physics Research Centre, Department of Physics, Jadavpur University, Kolkata - 700032, India}

\author{Somnath Kar}
\affiliation{Nuclear and Particle Physics Research Centre, Department of Physics, Jadavpur University, Kolkata - 700032, India}

\author{Argha Deb}
\affiliation{Nuclear and Particle Physics Research Centre, Department of Physics, Jadavpur University, Kolkata - 700032, India}
\affiliation{School of Studies in Environmental Radiation \& Archaeological Sciences, Jadavpur University, Kolkata - 700032, India}

\author{Premomoy Ghosh}
\affiliation{Variable Energy Cyclotron Centre, HBNI, 1/AF Bidhan Nagar, Kolkata - 700064, India}

\date{\today}

\begin{abstract}
Charged-particle multiplicity and summed values of the transverse momentum ($p_{\rm T}$) have been utilized for estimating
forward-backward (FB) correlation strength for EPOS3 simulated proton-proton ($pp$) events with and without hydrodynamical
evolution of particles at center-of-mass energies $\sqrt{s}$  = 0.9, 2.76, and 7 TeV for different pseudorapidity window width ($\delta\eta$)
and gap ($\eta_{gap}$) between the FB windows. We have studied the variation of FB correlation strength with $\eta_{gap}$, 
$\delta\eta$, $\sqrt{s}$, $p_{\rm T}$ cuts and multiplicity classes.
Results are compared with the corresponding ALICE and ATLAS data. EPOS3 model qualitatively reproduces the overall variation of correlation strength 
of the LHC data. However, quantitative agreement is better for $pp$ events, generated using EPOS3 with
hydrodynamical evolution of particles, with ATLAS data.
\end{abstract}


\maketitle

\section{Introduction}
In ultrarelativistic high-energy collisions, the study of correlations between produced particles in different pseudorapidity ($\eta$) regions gives us an opportunity to understand the dynamics of multiparticle interactions and their hadronization. In general these correlations are of two types: short-range correlations (SRCs) and long-range correlations (LRCs)~\cite{ref1, UA5_3, ref12, ref5}. Particles with lower transverse momentum ($p_{\rm T}$) are generated via soft processes~\cite{ref3} and are believed to be correlated  weakly over large $\eta$ range (LRC). The particles in the high-$p_{\rm T}$ regime, which are produced via harder perturbative processes, are strongly correlated over short pseudorapidity distances (SRC)~\cite{ref6}. With the gradual increase of particle momentum from soft regime to hard, the correlations strength is found to be weakened over large $\eta$ separations~\cite{ref4, ref43}. In different experiments and theoretical models, short-range correlations are considered to be localized over $|\eta|$ $\sim$ 1 units of pseudorapidity whereas long-range correlations extend over a wider range of pseudorapidity ($|\eta| > $ 1)~\cite{ref11}.

Forward-backward (FB) correlation, a robust tool to explore both the SRC and the LRC, plays important role in understanding initial state fluctuations in different collision systems like hadronic or nuclear. Pairs of pseudorapidity intervals equal in size and symmetrically located in the forward (beam direction) and backward (opposite to the beam direction) direction with respect to the collision vertex are considered as forward and backward windows, respectively. Event-by-event variations of different observables in FB windows can be used to construct FB correlation coefficients~\cite{UA5_3, refSLLim, ref6}. 

Several experimental studies on FB correlations had been previously carried out for different collision systems including electron-positron ($e^{+}e^{-}$), proton-proton ($pp$), proton-antiproton ($p\bar {p}$), proton-nucleus ($pA$), and nucleus-nucleus ($AA$)~\cite{ref1, ref12, UA5_3, ref4, ref7, ref8, ref9, ref10, ref13, ref14, ref15, ref16, ref17}. Though, there was no FB multiplicity correlation found in $e^{+}e^{-}$ annihilation~\cite{ref9}, but in hadronic collisions ($pp$/$p\bar {p}$) or in heavy-ion collisions with higher energies at the Super Proton Synchrotron (SPS)~\cite{ref1, ref12, UA5_3}, the Tevatron~\cite{ref13}, the Relativistic Heavy Ion Collider (RHIC)~\cite{ref14, ref15}, and the Large Hadron Collider (LHC)~\cite{ref4, ref16, ref17}, a considerable correlation strength was observed. All these experimental observations offer a cornucopia of scopes to testify various theoretical and/or phenomenological models for a possible explanation of the FB correlation exploiting different correlation coefficients between the multiplicities ($n-n$), the transverse momentums ($p_{T}-p_{T}$) and the transverse momentum and the multiplicity of charged particles ($p_{T}-n$).

Incipiently, the Dual Parton Model (DPM)~\cite{ref5, dpmFBcorr} and the Quark Gluon String Model (QGSM)~\cite{ref18} came up with the prediction of the possible long-range correlations taking into account the multiple parton-parton interactions. The Monte Carlo version of the QGSM~\cite{ref20}, which successfully described ALICE data in terms of FB correlation, showed that the superposition of  different multistring processes with different mean multiplicities in $pp$ collisions at various center-of-mass energies could be the source of FB correlations strength. The String Fusion Model (SFM)~\cite{SFM} investigated the long-range correlations with the idea of possible interactions between strings, highlighting different types of FB correlations as mentioned above~\cite{SFM1}. Furthermore, the Monte Carlo version of SFM predicted and reproduced the LHC data reasonably well in hadronic and nuclear collisions~\cite{MCSFM, MCSFM1}. The FB correlations were also studied via string percolation mechanism in $pp$ collisions~\cite{percolFB}. The study of FB correlations in the Color Glass Condensate model (CGC)~\cite{refCGC, ref21, CGC_LRC, ref21_CGC} showed that the initial state correlations and density fluctuations could lead to the observed long-range correlations among final-state particles foreseeing the centrality-dependent growth of LRC in heavy-ion collisions~\cite{ref15}.

Recent studies on high-multiplicity $pp$ and $pPb$ collisions at the LHC and $dAu$ collisions at the RHIC exhibit unforeseen features of collectivity~\cite{ref22, ref23, ref24, ref25, ref26, ref27, ref28}. Although hydrodynamical modeling remains a successful description to the properties of  medium produced in heavy-ion collisions, recently such approach is found to be applicable in small systems ($pp$ and/or $pPb$) at the LHC energies. The EPOS3 model with in-build hydro feature~\cite{ref29, ref30} remains successful in describing ALICE data~\cite{ref31} for the charged-particle flow and shows some hint of long-range ridge-like structure in high-multiplicity $pp$ collisions at $\sqrt{s}$ = 7 and 13 TeV~\cite{ref32} and $pPb$ collisions at $\sqrt{s_{\rm NN}}$ = 5.02 TeV~\cite{ref30, ref33}. ATLAS experiment at the LHC shows that EPOS simulation underestimates the FB correlations strength for $pp$ collisions at 13 TeV~\cite{ref16}, though the hydro feature of EPOS model remained unexplored and was not tested for all the available energies  at the LHC. Therefore, it is worth mentioning that the physics behind the FB correlation remains inconclusive even after different experimental and theoretical attempts and recent developments on the high-multiplicity events of small systems ($pp$/$pA$), which resemble many heavy-ion outcomes, demand further studies in this direction.

In this work, we, therefore, have used EPOS3 simulation code with and without hydrodynamical evolution of particles (referred as 
``with and without hydro" in rest of the texts) to explain the measured FB correlations in several rapidity window in $pp$ collisions at 
$\sqrt{s}$  = 0.9, 2.76, and 7 TeV. We have reported the multiplicity and summed transverse momentum FB correlations for the charged 
particles using different kinematics to comply with the experimental measurements. 

This paper is organized as follows: the formulation of FB correlation coefficients is mentioned in Sec.~\ref{sec2}. Section~\ref{sec3} 
discusses briefly about EPOS3 event generator and simulated events. Selection of EPOS3 generated events and different FB windows 
following ALICE and ATLAS kinematics is described in Sec.~\ref{sec4}. In Sec.~\ref{sec5}, the dependences of multiplicity and summed-$p_{T}$
FB correlation coefficients on the separation of pseudorapidity windows ($\eta_{gap}$), the width of the pseudorapidity window 
($\delta\eta$), the collision energy ($\sqrt{s}$), the minimum transverse momentum ($p_{\rm T_{min}}$), and the charged-particle multiplicity
have been presented in detail and compared with corresponding ALICE~\cite{ref17} and ATLAS~\cite{ref4} data. Finally, the paper ends with summary and conclusions 
in Sec.~\ref{sec6}.
\section{Forward-backward charged-particle correlation coefficient}\label{sec2}
In general, FB correlations between produced particles can be categorized into three main types~\cite{refFBtypes1}: 
\begin{itemize}
\item{$n - n$, the correlation between charged particles multiplicities} 
\item{$p_{T}-p_{T}$, the correlation between mean or summed transverse momenta of charged particles}
\item{$p_{T}-n$, the correlation between mean or summed transverse momenta in one pseudorapidity interval and the multiplicity of charged particles in another pseudorapidity interval}
\end{itemize}
The FB correlation strength is measured in a coordinate system with origin $\eta$ = 0 which is always located 
at midrapidity, i.e., the collision vertex.
\begin{figure}
\centering
\includegraphics[scale=0.35,keepaspectratio]{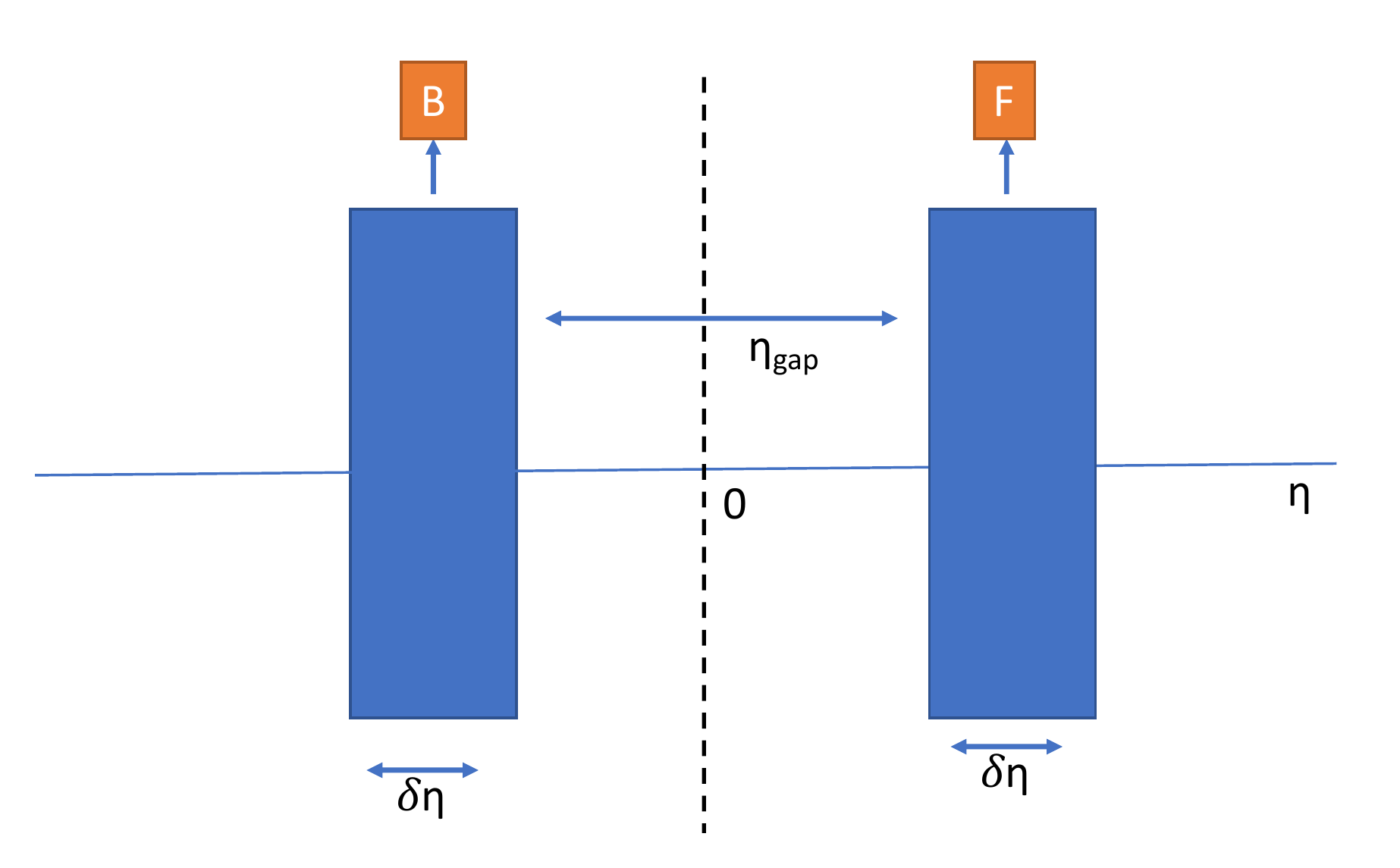}
\caption{(Color online) Construction of forward (F) and backward (B) window.}
\label{fig1}
\end{figure}
Two pseudorapidity intervals are selected, one in the forward ($\eta >$ 0) and another in the backward hemispheres ($\eta < $0) 
in the center-of-mass system. Figure~\ref{fig1} shows forward and backward window construction where, $\eta_{gap}$ being the 
gap between the window pairs and $\delta\eta$ being the width of each window. The FB correlation strength can be obtained 
from a linear regression analysis of the average charged-particle multiplicity in the backward hemisphere ($\eta < $ 0), 
$\langle N_{b}\rangle_{N_{f}}$, as a function of the event multiplicity in the forward hemisphere ($\eta > $ 0), $N_{f}$, 
such that,
\begin{equation}
\langle N_{b}\rangle_{N_{f}} = a + b_{corr} (mult) N_{f}
\label{eq1}
\end{equation}
where $a$  is a constant and $b_{corr}$ (mult) measures the multiplicity correlation strength~\cite{{ref1, ref5}}. If linear relation of Eq.~(\ref{eq1}) 
holds, then $b_{corr}$ (mult) can be estimated using the following formula of Pearson correlation coefficient:
\begin{equation}
b_{corr} (mult) = \frac{\langle N_{f} N_{b}\rangle - \langle N_{f}\rangle \langle N_{b}\rangle}{\langle N_{f}^{2}\rangle - \langle N_{f}\rangle^{2}} = \frac{D^{2}_{bf}}{D^{2}_{ff}}
\label{eq2}
\end{equation}
In Eq.~(\ref{eq2}), $D_{bf}^{2}$ (covariance) and $D^{2}_{ff}$ (variance) are the backward-forward and forward-forward 
dispersions respectively~\cite{ref5, ref6}.

Since, the charged-particle multiplicity is an extensive quantity, the FB multiplicity correlation strength is affected by the 
so-called ``volume fluctuations" which originate from event-by-event fluctuations of the number of participating nucleons. 
To avoid such fluctuations, we can consider, intensive observables like the sum of the absolute transverse momentum of particles 
within the observation windows. Similar to the multiplicity correlation, forward-backward summed-$p_{\rm T}$ correlation 
coefficient can be extracted using the following formula:
\begin{equation}
\small b_{corr} (\Sigma p_{T}) = \frac{\langle \Sigma p_{T_{f}} \Sigma p_{T_{b}}\rangle -  \langle \Sigma p_{T_{f}}\rangle \langle\Sigma p_{T_{b}}\rangle}{\langle(\Sigma p_{T_{f}})^{2}\rangle -  \langle\Sigma p_{T_{f}}\rangle^{2}}
\label{eq3}
\end{equation}
Here, $\Sigma p_{T_{f}}$ and $\Sigma p_{T_{b}}$ are the event summed transverse momentum in forward and backward 
window, respectively.
\begin{figure*}[hbt!]
\centering
\includegraphics[scale=0.32,keepaspectratio]{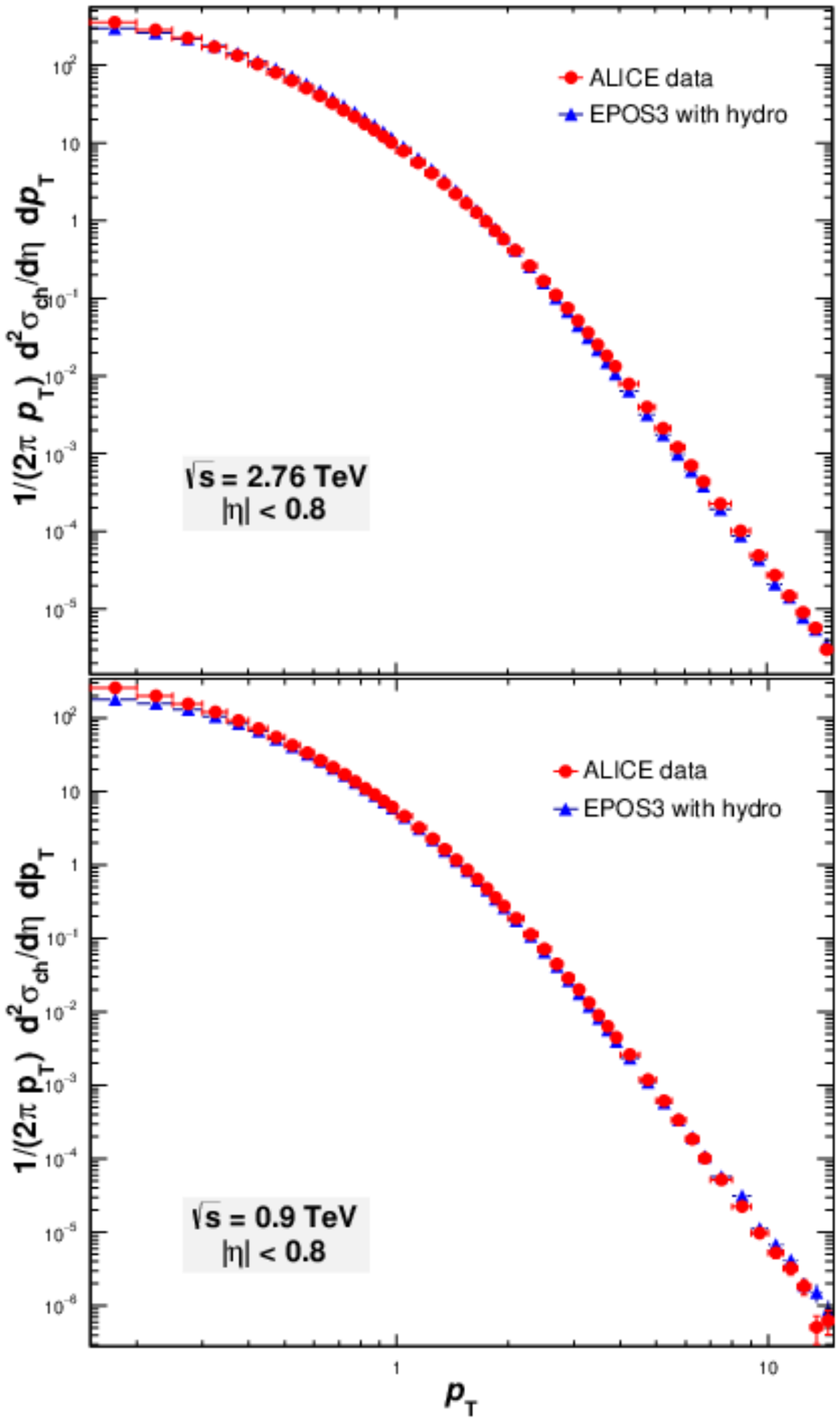}
\includegraphics[scale=0.32,keepaspectratio]{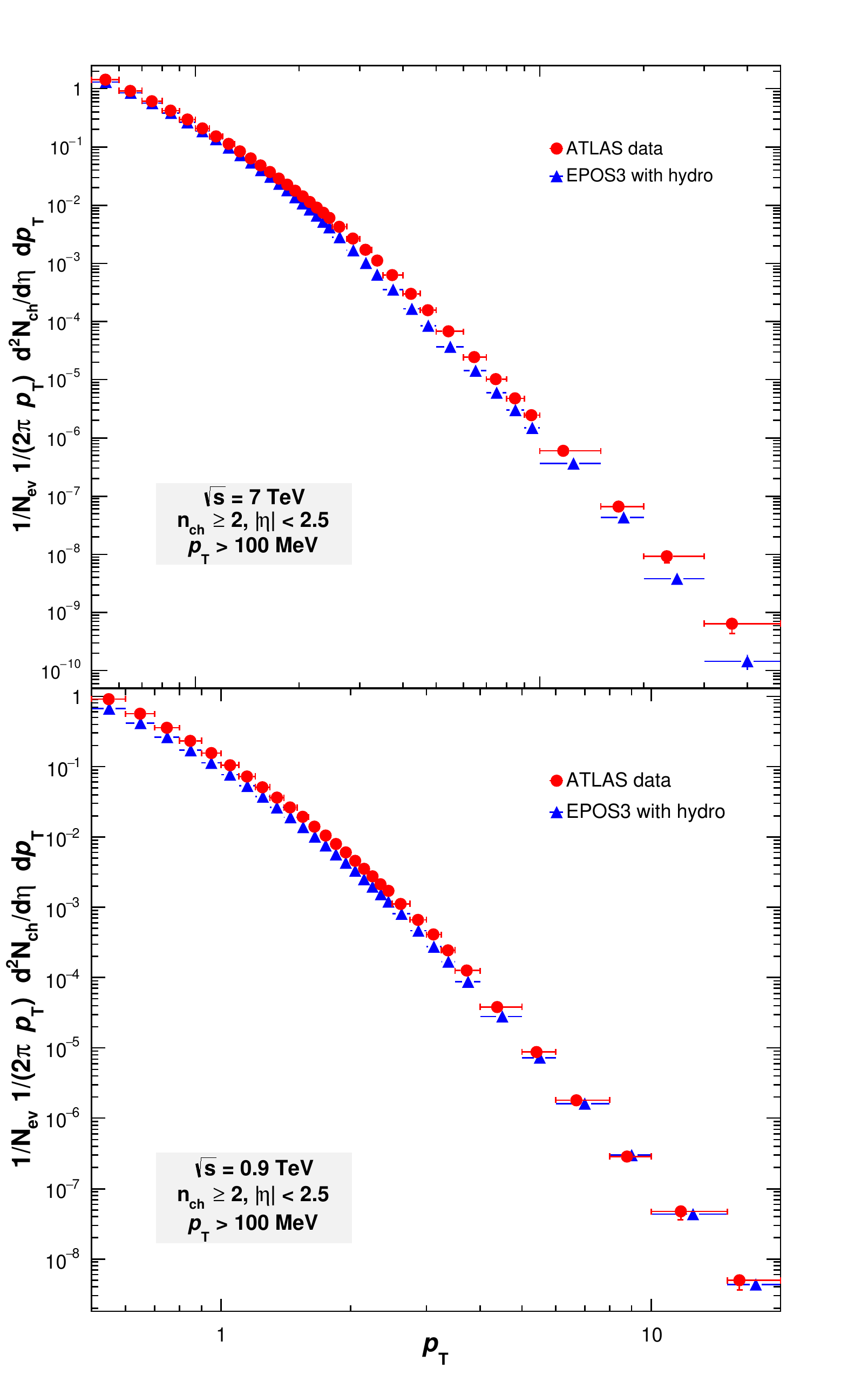} 
\caption{(Color online) (Left) Differential cross section of charged particle as a function of $p_{\rm T}$ from EPOS3 with hydro generated minimum-bias events in $pp$ collisions at $\sqrt{s}$ = 2.76 (upper panel) and 0.9 TeV (lower panel), compared to ALICE data~\cite{ref36}. (Right) Charged-particle multiplicities as a function of $p_{\rm T}$ from same EPOS3 events in $pp$ collisions at $\sqrt{s}$ = 7 TeV (upper panel) and 0.9 TeV (lower panel) compared to ATLAS data~\cite{ref37}.}
\label{fig2}
\end{figure*}

Similarly, the correlation strength between mean or summed transverse momenta and the charged-particle multiplicity can also be described following the formula of Pearson correlation coefficient. However, we have explored first two types of FB correlation in details in this paper.
%
\section{The EPOS3 Model}\label{sec3}
The $p$QCD-inspired hybrid Monte Carlo event generator EPOS3 uses Gribov-Regge multiple scattering framework for particle 
productions in high-energy collisions. The most unique feature of EPOS3 model is to use a common theoretical scheme for the 
particle production in $pp$, $pA$, and $AA$ collisions. Unlike many other Monte Carlo 
event generators, EPOS3 generates real event which does not introduce any “test particles” and all kinds of fluctuations are treated 
on the basis of event-by-event fluctuations~\cite{ref34}. 

In this approach an individual scattering is termed as a ``Pomeron". For a given pomeron, the corresponding chain of partons is treated 
as parton ladder which may be considered as a longitudinal color field or a flux tube, carrying transverse kinks from the initial hard 
scatterings~\cite{ref35}. In a collision, many elementary parton-parton hard scatterings form a large number of flux tubes that expand and 
are fragmented into string segments. Some of these flux tubes constitute the bulk matter or a medium which thermalizes and undergoes a 
three-dimensional (3D)+1 viscous hydrodynamical evolution and hadronizes via usual Cooper-Frye formalism at a ``hadronization temperature", T$_{\rm H}$. 
These segments form the so-called ``core" and this collective expansion takes place till soft hadrons (low $p_{\rm T}$ particles) freeze-out. 
Other string segments having high transverse momentum that are close to the surface leave the bulk matter and hadronize (including 
jet hadrons) via the Schwinger mechanism.Those segments form the so-called ``corona". Rest of the string segments which have enough 
energy to escape the bulk matter constitute the ``semihard" or intermediate-$p_{\rm T}$ particles. At the time of escaping, these segments 
may pick up quarks or antiquarks from the bulk matter inheriting the imprints of its properties. 

Using EPOS3 model, we generated 3 million minimum-bias $pp$ events for center-of-mass energies 0.9, 2.76, and 7 TeV,
for each of the options, with and without hydro. To validate the generated event samples of different center-of-mass energies, we compared EPOS3 simulated events with ALICE and ATLAS data. Figure~\ref{fig2} shows that the differential cross section 
of charged particles as a function of $p_{\rm T}$ as measured by ALICE experiment in $pp$ collisions at $\sqrt{s}$ = 0.9 and 2.76 TeV~\cite{ref36} 
and charged-particle multiplicities as a function of $p_{\rm T}$ by ATLAS experiment at $\sqrt{s}$ = 0.9 and 7 TeV~\cite{ref37} have been successfully
reproduced by the simulated events at the chosen energies.
\section{FB Window and Event Selection}\label{sec4}
We have studied FB correlations following ALICE~\cite{ref17} and ATLAS~\cite{ref4} kinematics.
\vspace*{-\baselineskip}
\subsection{ALICE kinematics}\label{sec4_1}
We have selected EPOS3 simulated events having a minimum of two charged particles in the kinematic interval 0.3 $< p_{T} <$ 1.5 
GeV and $|\eta| <$ 0.8 following ALICE~\cite{ref17} kinematics. We have divided the chosen pseudorapidity space into two windows about the 
collision center, i.e., $\eta$ = 0. One is forward window (F) ($\eta >$ 0) and another is backward window (B) ($\eta <$ 0). Two pseudorapidity 
intervals of equal width ($\delta\eta$) have been taken symmetrically from the F and B windows. Four different values of $\delta\eta$ are 
taken, i.e., $\delta\eta$ = 0.2, 0.4, 0.6 and 0.8. Also, we have considered three different values of $\eta_{gap}$ (the separation between the 
forward and backward pseudorapidity intervals), i.e., $\eta_{gap}$ = 0, 0.4 and 0.8. We have studied the forward-backward charged-particle 
multiplicity and summed-$p_{\rm T}$ correlations for each value of $\eta_{gap}$ considering possible values of $\delta\eta$.
\subsection{ATLAS kinematics}\label{sec4_2}
While following ATLAS kinematics~\cite{ref4}, EPOS3 generated events are chosen with a minimum of two charged particles with $p_{\rm T} >$ 0.1 GeV and $|\eta| <$ 2.5. Equal intervals in pseudorapidity of size $\delta\eta$ = 0.5 are chosen for all possible combinations of forward ($\eta >$ 0) and backward ($\eta <$ 0) windows with equal or different $\eta_{gap}$.
\section{Results and Discussions}\label{sec5}
\subsection{Multiplicity correlation}
The correlation between the forward and backward multiplicities $N_{f}$ and $N_{b}$ of the produced charged particles 
has been extensively studied for EPOS3 generated $pp$ events at three center-of-mass energies $\sqrt{s}$ = 
0.9, 2.76, and 7 TeV and compared with corresponding experimental data.
\subsubsection{Analysis considering ALICE kinematics}
\begin{figure}
\centering
\includegraphics[scale=1.05,keepaspectratio]{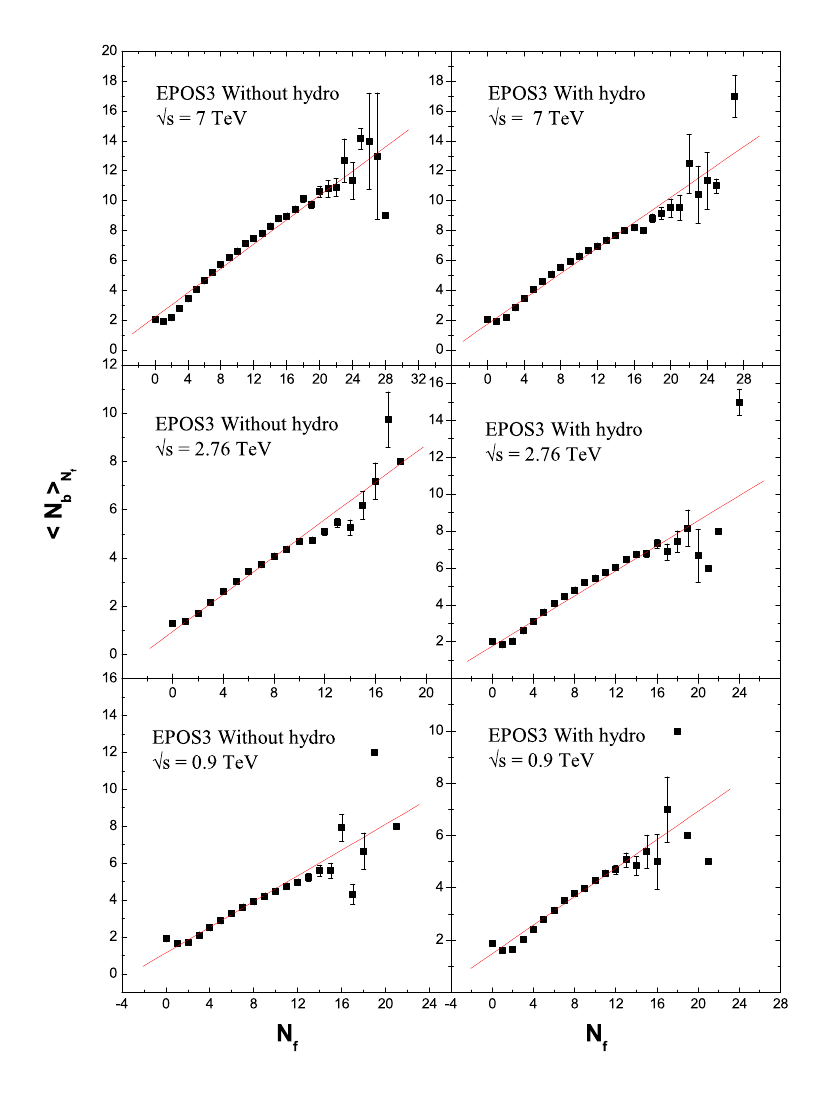}
\caption{(Color online) Variation of $\langle N_{b}\rangle_{N_{f}}$ with $N_{f}$ for FB window width $\delta\eta$ = 0.6 and $\eta_{gap}$ = 0.4 for 
EPOS3 generated $pp$ events with (right panel) and without (left panel) hydro at three center-of-mass energies $\sqrt{s}$ = 0.9, 2.76, and 7 TeV.}
\label{fig5}
\end{figure}

We have performed the following analysis considering events and FB windows as described in Sec.~\ref{sec4_1}. 
Figure~\ref{fig5} shows the dependence of the average charged-particle multiplicity in the backward window 
($\langle N_{b}\rangle_{N_{f}}$) on the charged-particle multiplicity ($N_{f}$) in the forward window taking 
window width $\delta\eta$ = 0.6 and $\eta_{gap}$ = 0.4 for EPOS3 simulated $pp$ events with and without 
hydro at $\sqrt{s}$ = 0.9, 2.76, and 7 TeV.
We found a linear correlation between $\langle N_{b}\rangle_{N_{f}}$ and $N_{f}$ as depicted in Eq.~(\ref{eq1}).
The data points are well fitted by a linear function, shown by the red lines in all panels in Fig.~\ref{fig5}. Henceforth, we have used 
Pearson correlation coefficient of Eq.~(\ref{eq2}) for the calculation of multiplicity correlation strength, $b_{corr}$(mult), and performed 
the following studies:
\begin{figure}
\centering
\includegraphics[scale=1.0,keepaspectratio]{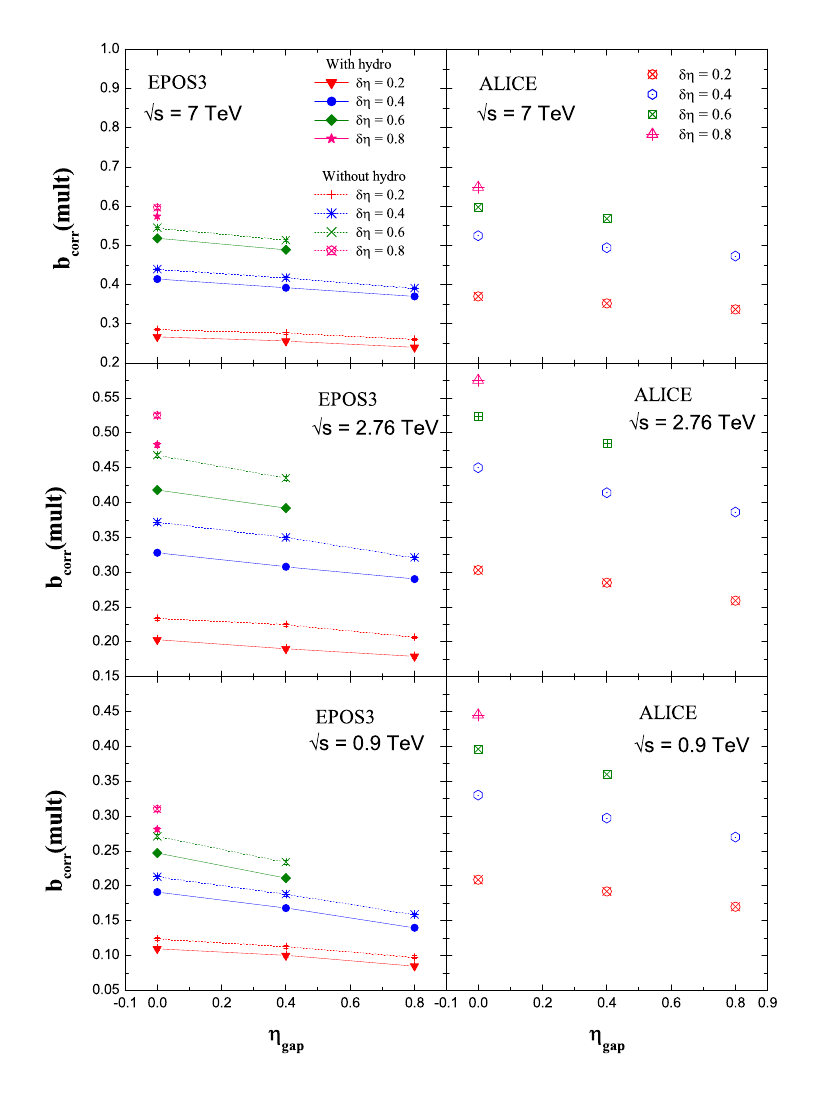}
\caption{(Color online) FB multiplicity correlation strength, $b_{corr}$ as a function of $\eta_{gap}$ for
$\delta\eta$ = 0.2, 0.4, 0.6 and 0.8 in $pp$ events at $\sqrt{s}$ = 0.9, 2.76, and 7 TeV. The left panel is for EPOS3 
generated $pp$ events considering with and without hydro and the right panel exhibits ALICE data~\cite{{ref17}}.}
\label{fig6}
\end{figure}
\begin{itemize}
\item{{\bf Dependence on the gap between FB windows ($\eta_{gap}$}):

The FB multiplicity correlation coefficient $b_{corr}$ as a function of $\eta_{gap}$ for four different window widths 
(as discussed in~\ref{sec4_1}) has been shown in Fig.~\ref{fig6} for the three collision energies $\sqrt{s}$ = 0.9, 2.76, and 7 TeV 
for EPOS3 simulated $pp$ events considering both with and without hydro (left panel). Right panel of Fig.~\ref{fig6} represents 
ALICE data~\cite{ref17}. It has been observed that $b_{corr}$ values for each center-of-mass energy decrease slowly with 
the increase of the gap between FB windows ($\eta_{gap}$). It is evident that the experimental values are higher than that of 
simulated values but the trend of dependence on $\eta_{gap}$ is in agreement with the experiment.
}
\item{{\bf Dependence on the width of FB windows ($\delta\eta$)}:

It can be seen from Fig.~\ref{fig6} that for a fixed separation between FB windows, $b_{corr}$ increases with 
the increase of window width ($\delta\eta$). For studying the nature of increase, $b_{corr}$ is plotted for most central 
window with respect to $\delta\eta$ in Fig.~\ref{fig7}. It shows that multiplicity correlation increases nonlinearly 
with window width $\delta\eta$. This dependence is in qualitative agreement with ALICE data~\cite{ref17}. The nonlinear 
dependence of $b_{corr}$ on $\delta\eta$ has been explained in a simple model reported by ALICE collaboration~\cite{ref17}, 
along with other approaches mentioned in~\cite{ref6, ref20, ref38, ref39}. The similar trend for both with and without hydro shows 
that the hydrodynamical evolution of the bulk matter has negligible effect on $b_{corr}$ as the SRC may be dominated due 
to event-by-event multiplicity fluctuations.
\begin{figure}
\centering
\includegraphics[scale=0.70,keepaspectratio]{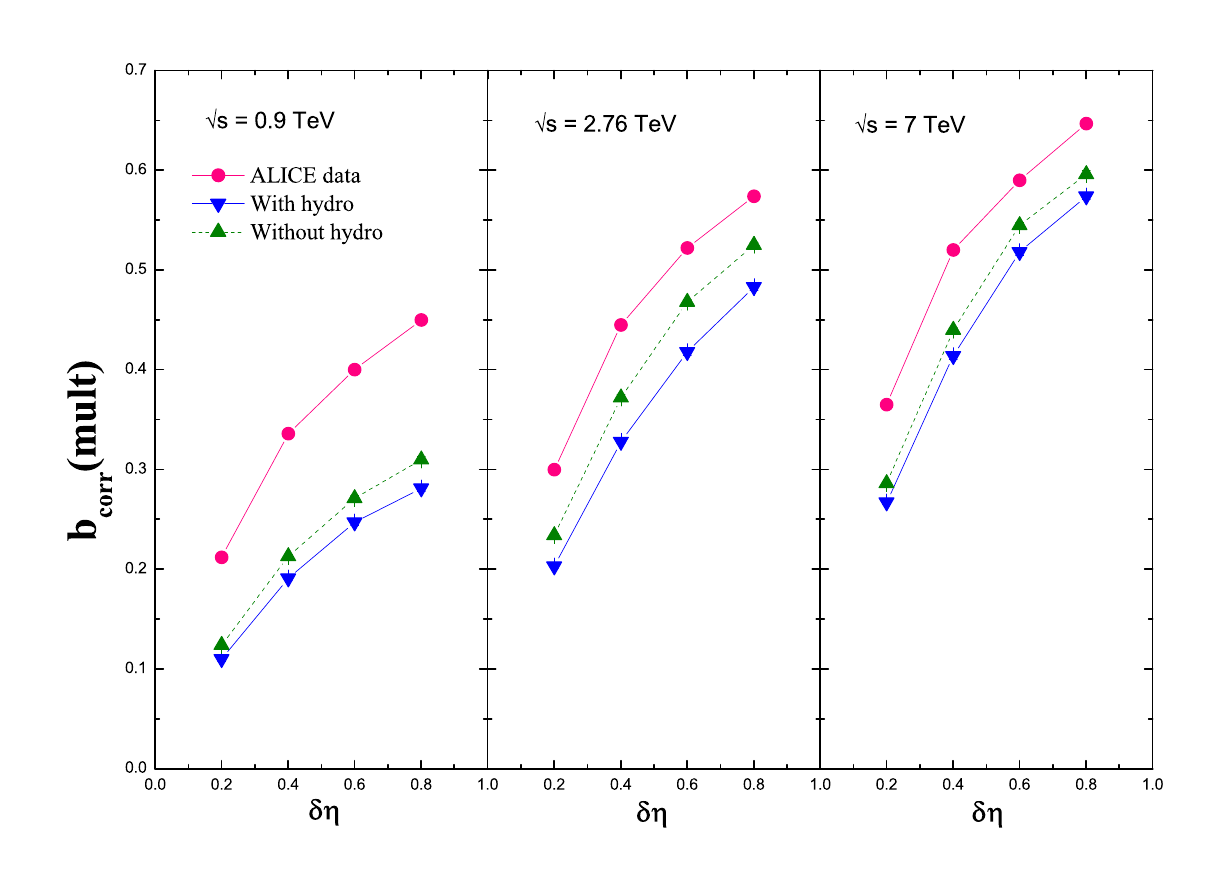}
\caption{(Color online) FB multiplicity correlation strength, $b_{corr}$ as a function of $\delta\eta$ for $\eta_{gap}$ = 0 in $pp$ 
collisions at $\sqrt{s}$ = 0.9, 2.76, and 7 TeV.}
\label{fig7}
\end{figure}
}
\item{{\bf Dependence on collision energy ($\sqrt{s}$)}:

It is evident from Fig.~\ref{fig6} and Fig.~\ref{fig7} that with the increase of collision energy FB multiplicity correlation 
increases. To have a closer look on energy dependence, the FB multiplicity correlation coefficient $b_{corr}$ has also 
been plotted with $\eta_{gap}$ for $\delta\eta$ = 0.4 at three center-of-mass energies in Fig.~\ref{fig8}. Although the 
slopes of the $\eta_{gap}$ dependence of $b_{corr}$ for three center-of-mass energies remain approximately constant 
for experimental data~\cite{ref17} as well as simulated events, it has been observed that the pedestal values of $b_{corr}$ 
increase with collision energy.
\begin{figure}
\centering
\includegraphics[scale=0.70,keepaspectratio]{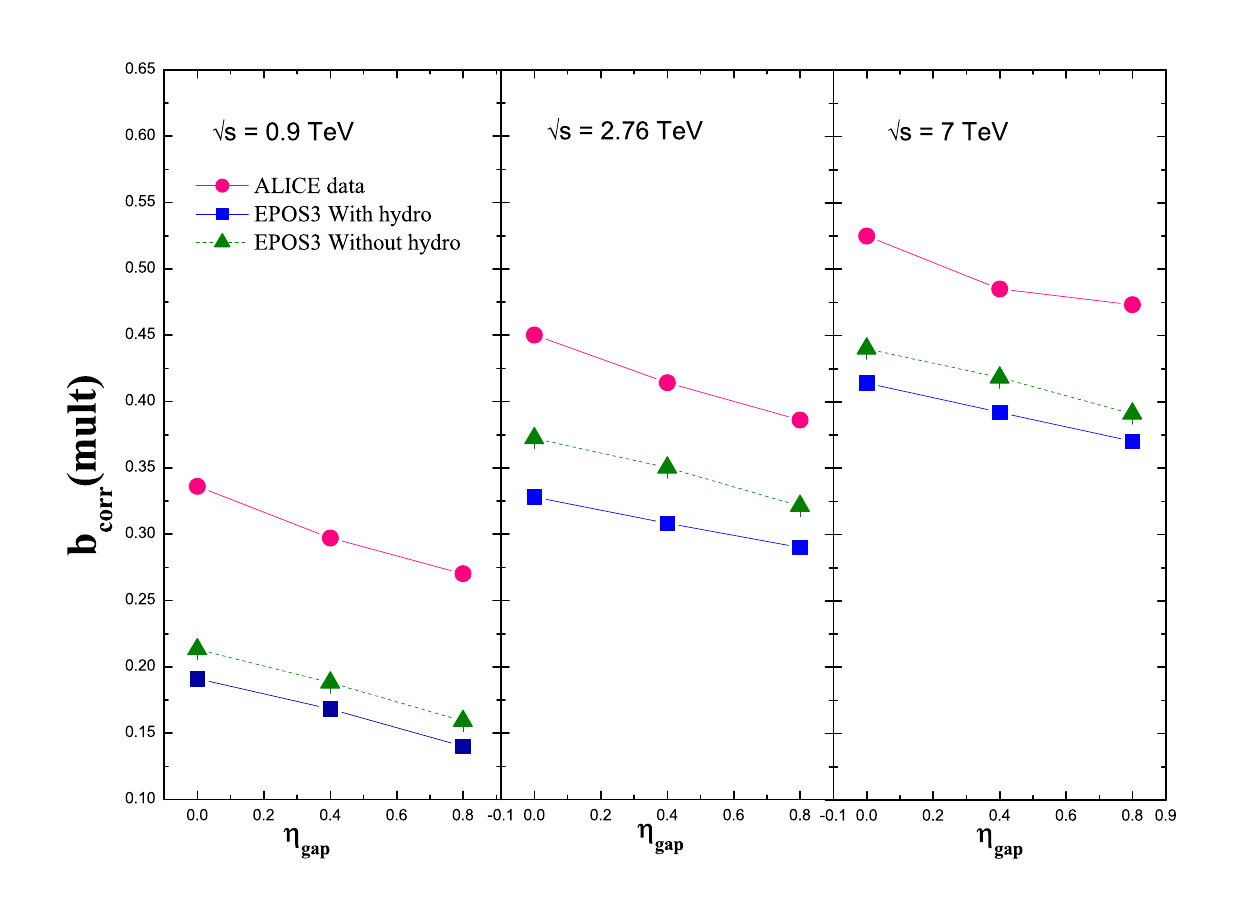}
\caption{(Color online) FB multiplicity correlation strength, $b_{corr}$ as a function of $\eta_{gap}$ for $\delta\eta$ = 0.4 in $pp$ 
collisions at $\sqrt{s}$ = 0.9, 2.76, and 7 TeV.}
\label{fig8}
\end{figure}
One of the reasons of this increase of the pedestal values of $b_{corr}$ with center-of-mass energy is the increase in 
mean multiplicity, $\langle N_{f}\rangle$. However, ALICE collaboration~\cite{ref17} has reported that if one chooses 
window sizes such that the mean multiplicities stays constant at different energies, the increase is still noticed. A strong 
energy dependence of $b_{corr}$ values were also reported by the UA5 collaboration~\cite{ref12} and ATLAS collaboration~\cite{ref4}.
}
\end{itemize}
\subsubsection{Analysis considering ATLAS kinematics}
\begin{figure*}
\centering
\includegraphics[scale=0.32,keepaspectratio]{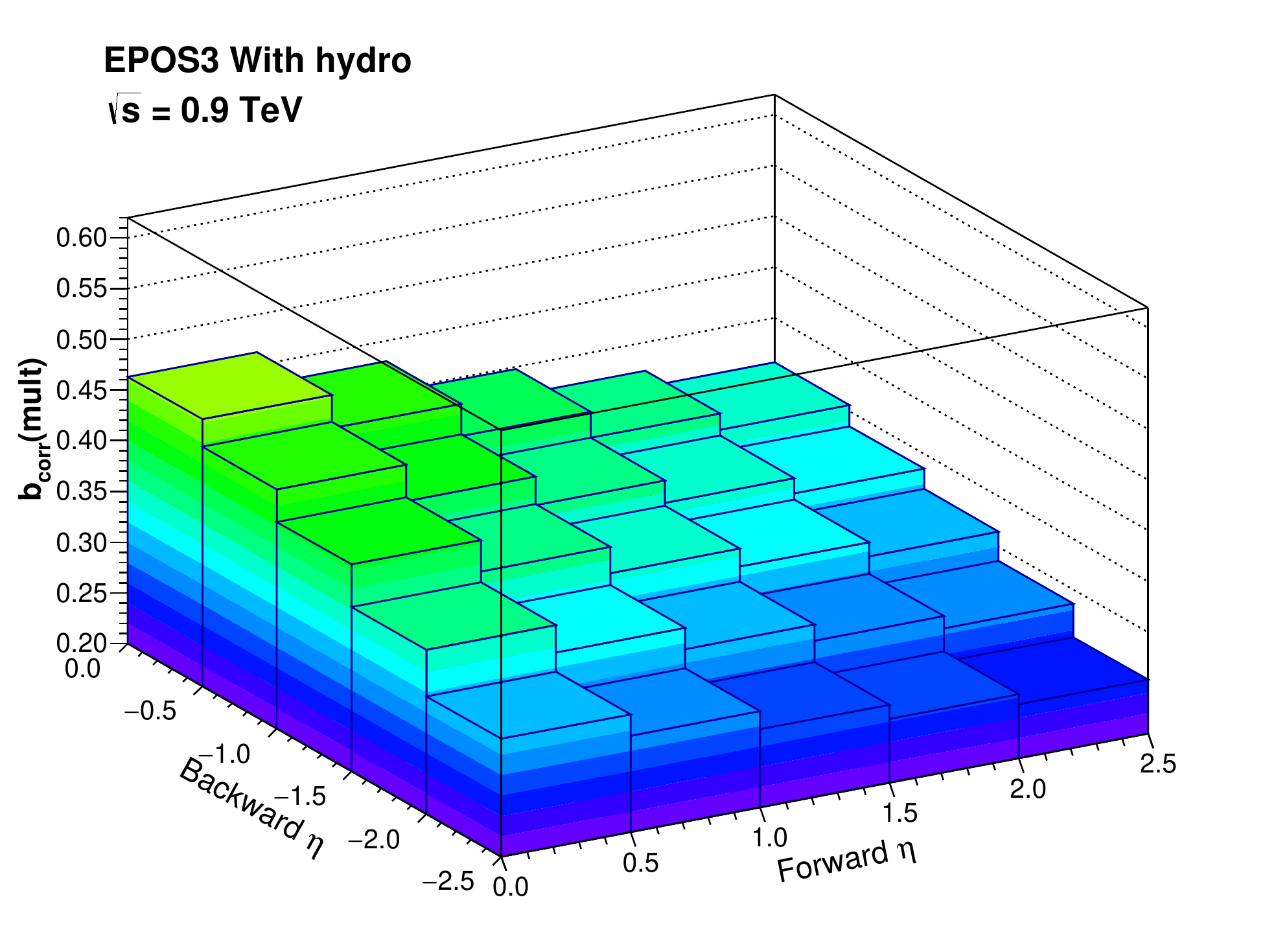} 
\includegraphics[scale=0.32,keepaspectratio]{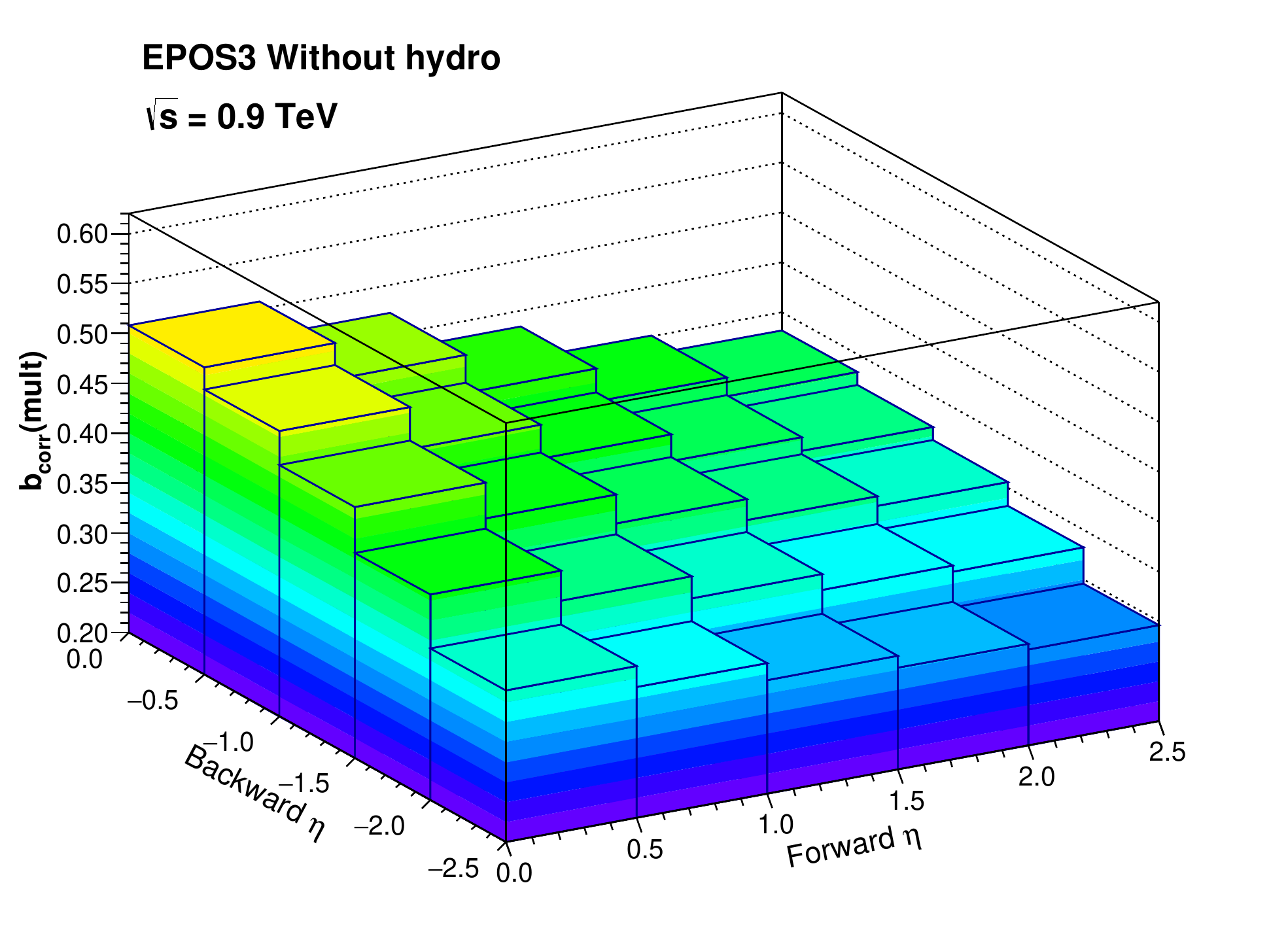}
\includegraphics[scale=0.32,keepaspectratio]{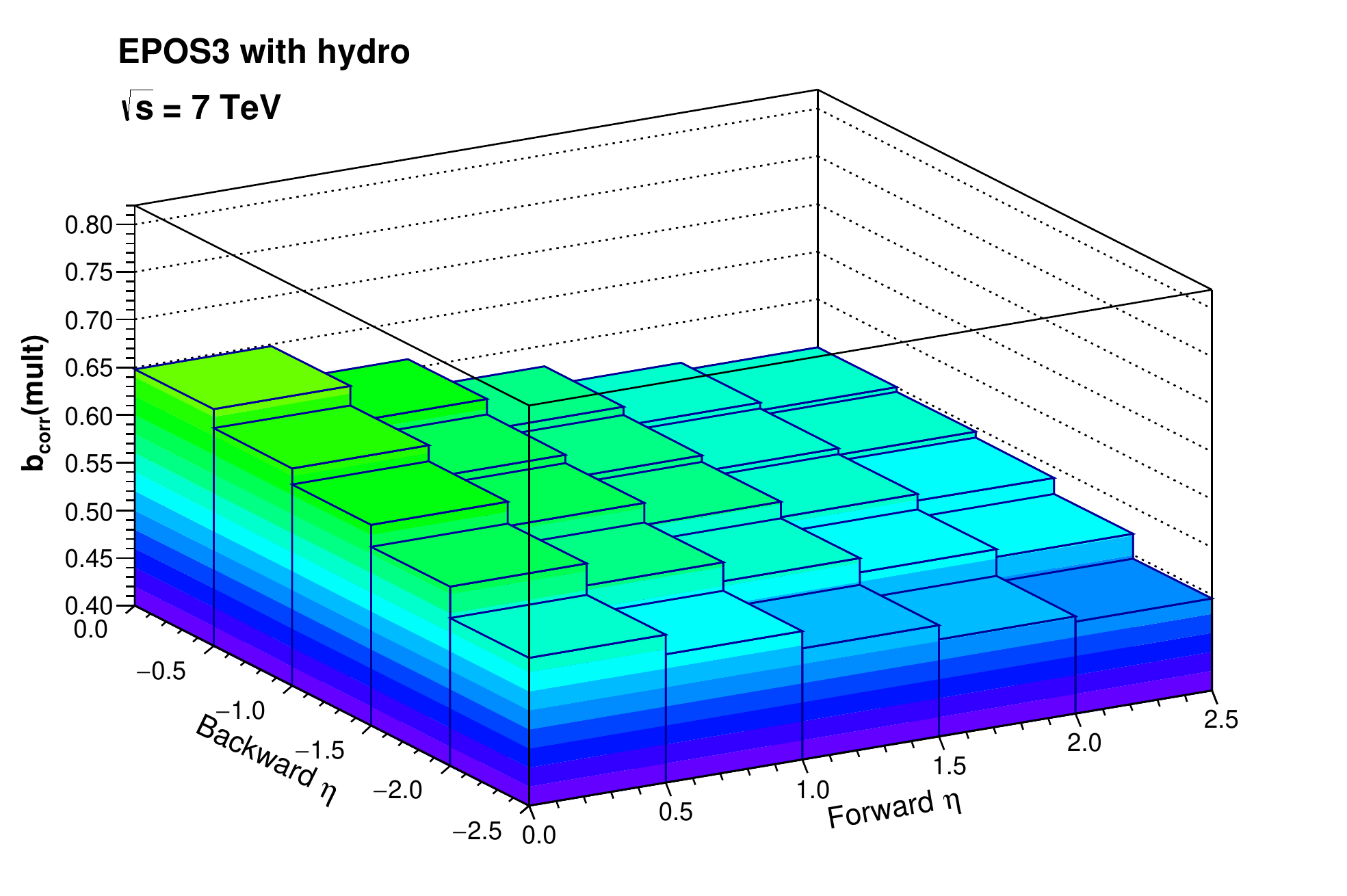} 
\includegraphics[scale=0.32,keepaspectratio]{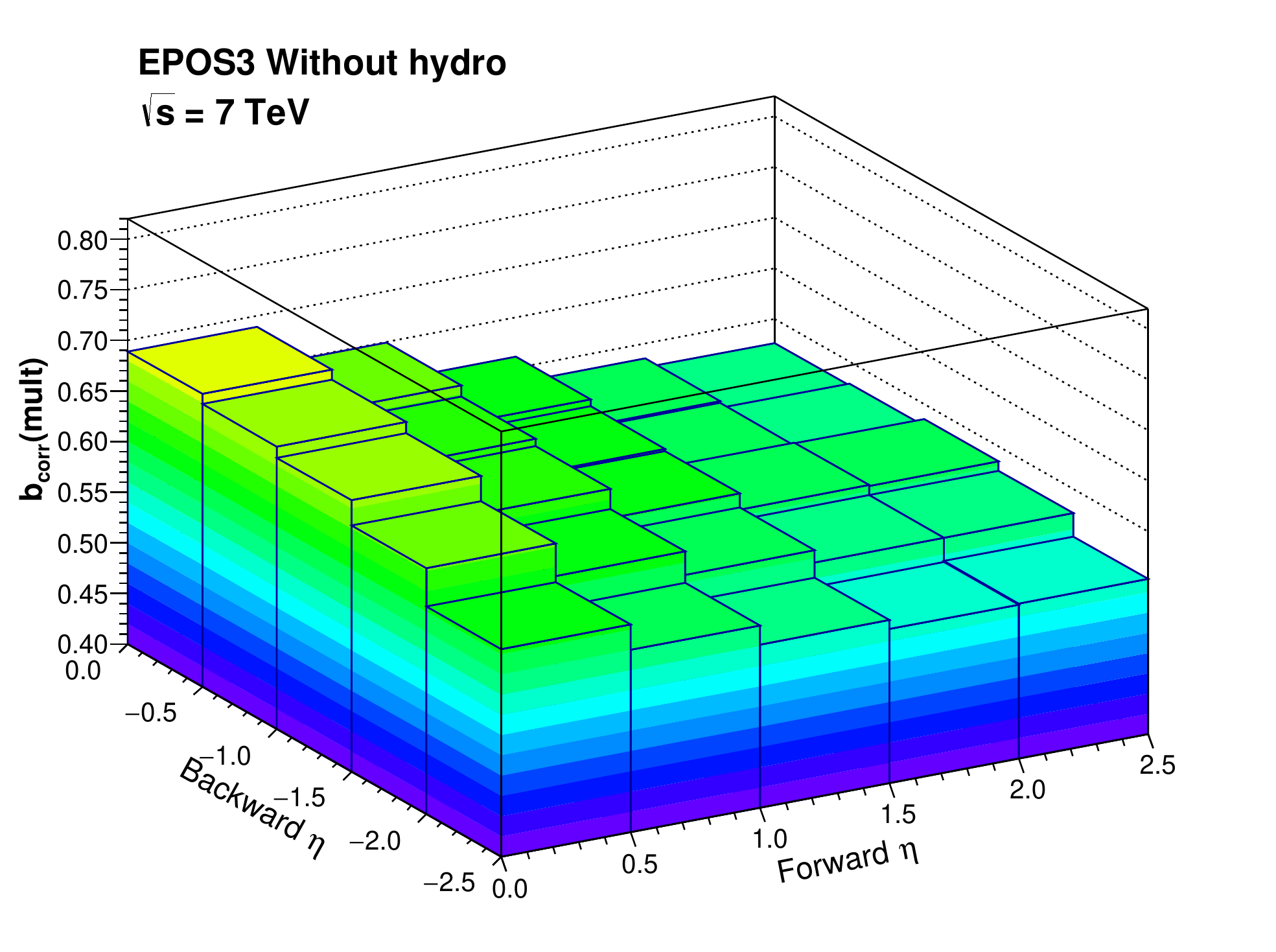} 
\caption{(Color online) Multiplicity correlation in a matrix of forward/backward $\eta$ intervals for $|\eta| <$ 2.5 and $p_{\rm T} >$ 0.1 
GeV for EPOS3 simulated events with at least two charged particles: (top-left) with hydro at 0.9 TeV, (top-right) without 
hydro at 0.9 TeV, (bottom-left) with hydro at 7 TeV, (bottom-right) without hydro at 7 TeV.}
\label{fig9}
\end{figure*}

We have done the following analyses considering events and FB windows as described in Sec.~\ref{sec4_2}.
The FB multiplicity correlations using EPOS3 simulated $pp$ events with and without hydro at $\sqrt{s}$ = 0.9 and 7 TeV
have been calculated for the full matrix of FB windows of width $\delta\eta$ = 0.5 as illustrated in Fig.~\ref{fig9} covering 
the whole range of pseudorapidity, $|\eta| < $ 2.5 and $p_{\rm T} >$ 0.1 GeV.
The main diagonal of Fig.~\ref{fig9} represents the symmetric FB windows with increasing separation. 
It is evident that the FB multiplicity correlation varies strongly with the $\eta_{gap}$s but weakly with the 
mean-$\eta$ value for a given separation for both with and without hydro in EPOS3 simulated $pp$ events 
at $\sqrt{s}$ = 0.9 and 7 TeV.
\begin{itemize}
\item {{\bf Dependence on the gap between FB windows ($\eta_{gap})$}:

\begin{figure}
\centering
\includegraphics[scale=0.42,keepaspectratio]{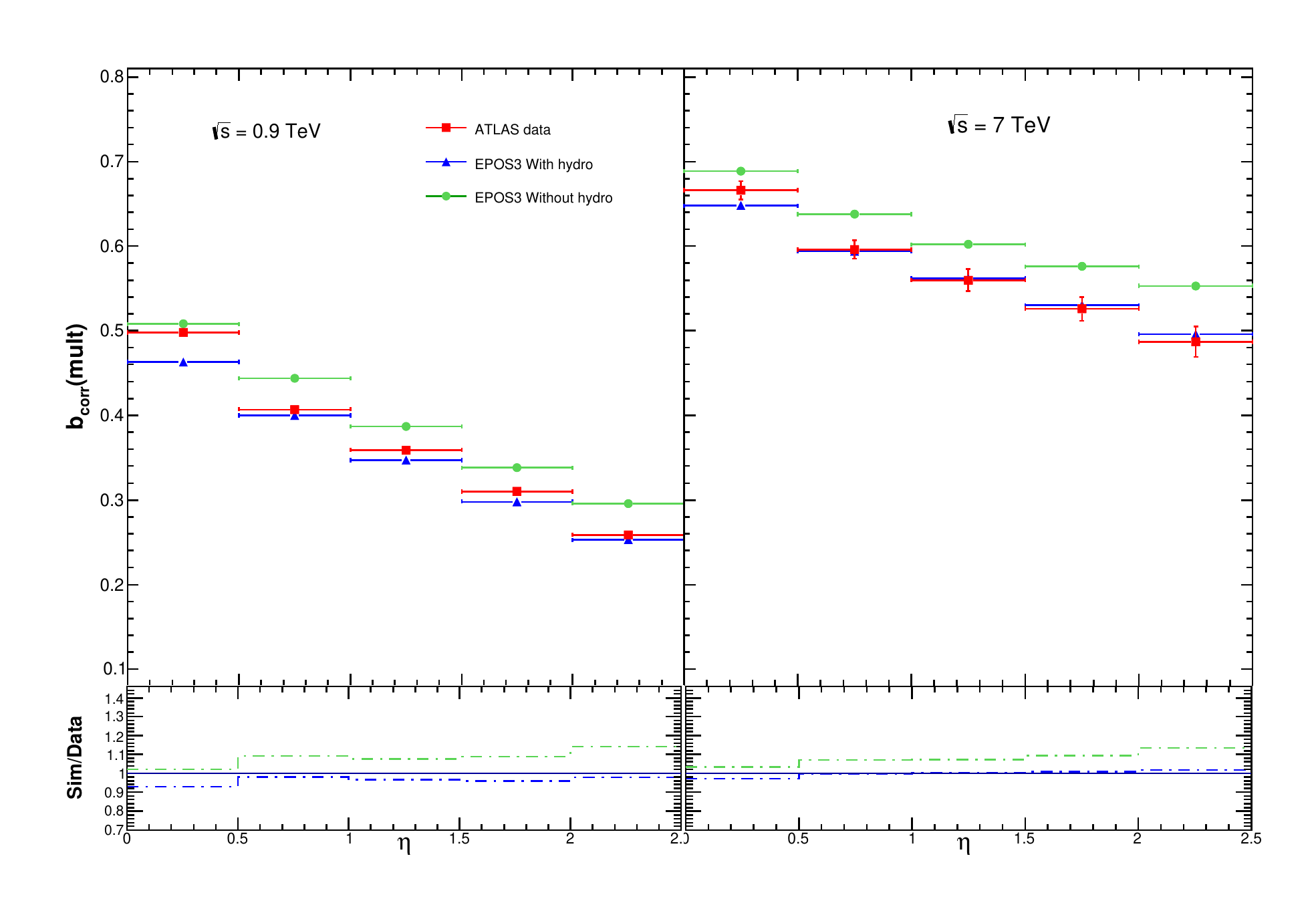}
\caption{(Color online) (Upper panel) FB multiplicity correlation in symmetrically opposite $\eta$ intervals for $pp$ events with 
$p_{\rm T} >$ 0.1 GeV and $|\eta| <$ 2.5. EPOS3 simulated events with and without hydrodynamics, compared to ATLAS 
data at $\sqrt{s}$ = 0.9 and 7 TeV. (Lower panel) Ratio of simulated events to ATLAS data.}
\label{fig10}
\end{figure}
The correlations between symmetrically opposite FB $\eta$-windows of equal width $\delta\eta$ = 0.5 have also been 
observed separately in Fig.~\ref{fig10} and compared to ATLAS results~\cite{ref4}. The lower panel represents 
the ratio between the simulated and experimental values for both the energies. It is interesting to note that the general trend 
is well reproduced by both type of EPOS3 simulated events. EPOS3 simulated events with hydro quantitatively reproduce the experimental
data for different $\eta_{gap}$s except the most central one at $\sqrt{s}$ = 7 TeV but underestimate the correlation 
strength at $\sqrt{s}$ = 0.9 TeV, whereas, events without hydro overestimate the same for both the energies.
}
\item {{\bf Dependence on center-of-mass energy ($\sqrt{s}$)}:

Fig.~\ref{fig11} represents the ratio of the above FB multiplicity correlation at $\sqrt{s}$ = 0.9 and 7 TeV for the simulated events as well as 
for the experimental data~\cite{ref4}. It has been found that the FB multiplicity correlation is higher for 7 TeV than 0.9 TeV and 
the relative difference is greater for the higher pseudorapidity gaps. Here, we can infer that similar to the data, in EPOS3 simulated 
events the LRC dominates over the SRC as the collision energy increases.
\begin{figure}
\centering
\includegraphics[scale=0.32,keepaspectratio]{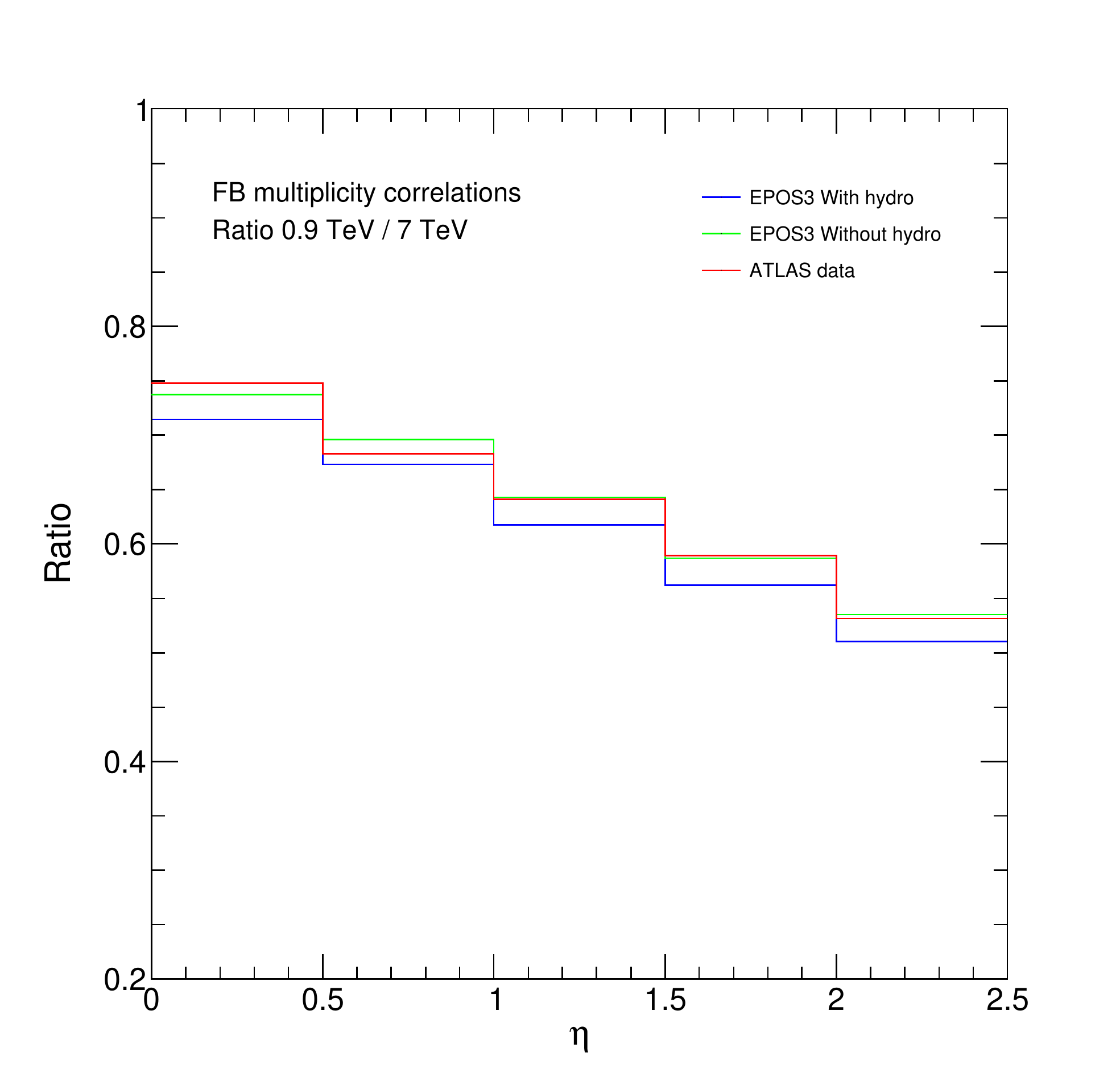}
\caption{(Color online) Ratio of the 0.9 TeV results to the 7 TeV results for EPOS3 with and without hydro and ATLAS data.}
\label{fig11}
\end{figure}
}
\item {{\bf Dependence on the minimum transverse momentum ($p_{\rm T_{min}}$)}

We know that in high-energy collisions with the increase of particle transverse momentum, there is a gradual transition
from soft processes to hard processes. To capture the contribution of this transition in multiplicity correlation, we have 
evaluated the value of $b_{corr}$ for seven different values of minimum transverse momentum ($p_{\rm T_{min}}$), i.e. 
$p_{\rm T_{min}}$ = 0.1, 0.2, 0.3, 0.5, 1.0, 1.5 and 2.0 GeV in case of symmetric FB windows with no separation for EPOS3 
simulated events with and without hydro for $pp$ collisions at $\sqrt{s}$ = 7 TeV and plotted in Fig.~\ref{fig12} along with the 
ATLAS data~\cite{ref4}. It has been found that the correlation decreases rapidly as $p_{\rm T_{min}}$ increases above a few 
hundred MeV following the same trend as in the experimental data. The decrease is more sharp for without hydro EPOS3 
events than with hydro. However, the agreement with experimental result is better for with hydro EPOS3 events.
\begin{figure}
\centering
\includegraphics[scale=0.65,keepaspectratio]{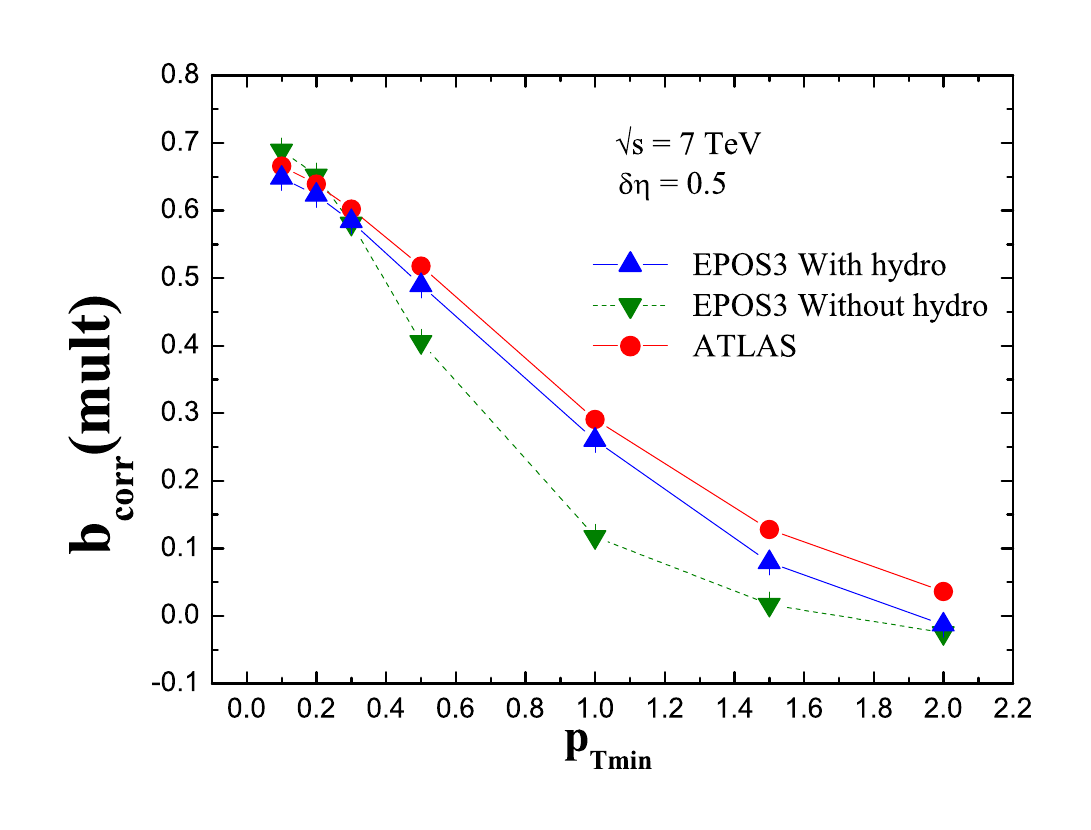}
\caption{(Color online) Forward-backward multiplicity correlations as a function of $p_{\rm T_{min}}$ for EPOS3 simulated events 
with and without hydro at $\sqrt{s}$ = 7 TeV compared with ATLAS data.}
\label{fig12}
\end{figure}
}
\end{itemize}
\subsection{Summed-$p_{\rm T}$ ($\Sigma p_{T}$) correlation}
The correlation among the summed values of the transverse momenta of the produced charged particles in 
forward and backward windows, $\Sigma p_{T_{_{f}}}$ and  $\Sigma p_{T_{_{b}}}$, has been studied for the same simulated 
events and compared with corresponding experimental data. We have estimated FB momentum correlation coefficient, 
$b_{corr} (\Sigma p_{T})$ using Eq.~(\ref{eq3}) and repeated the above analyses following ALICE~\cite{ref17} and ATLAS~\cite{ref4} kinematics.\\

\begin{figure}[hbt!]
\centering
\includegraphics[scale=0.74,keepaspectratio]{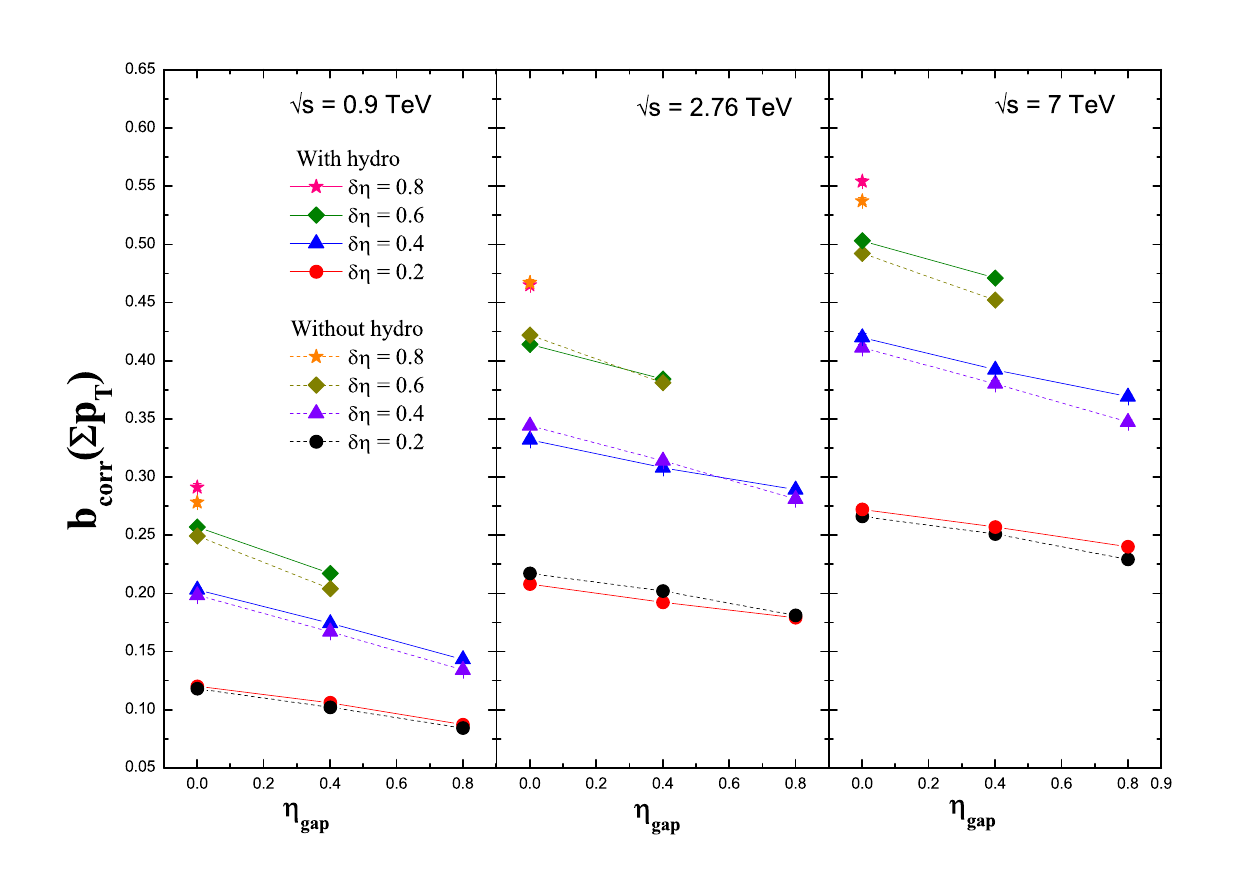}
\caption{(Color online) Forward-backward summed-$p_{\rm T}$ correlation as a function of $\eta_{gap}$ for four window widths $\delta\eta$ 
= 0.2, 0.4, 0.6 and 0.8 in EPOS3 generated $pp$ events at $\sqrt{s}$ = 0.9, 2.76, and 7 TeV.}
\label{fig13}
\end{figure}

Figure~\ref{fig13} transpires the fact that, similar to FB multiplicity correlation, FB momentum correlation strength also decreases 
gradually with the increasing gap between the FB windows ($\eta_{gap}$) for all window widths ($\delta\eta$) and maintains nearly 
constant slope. It increases with the increase of center-of-mass energy.

\begin{figure}[hbt!]
\centering
\includegraphics[scale=0.75,keepaspectratio]{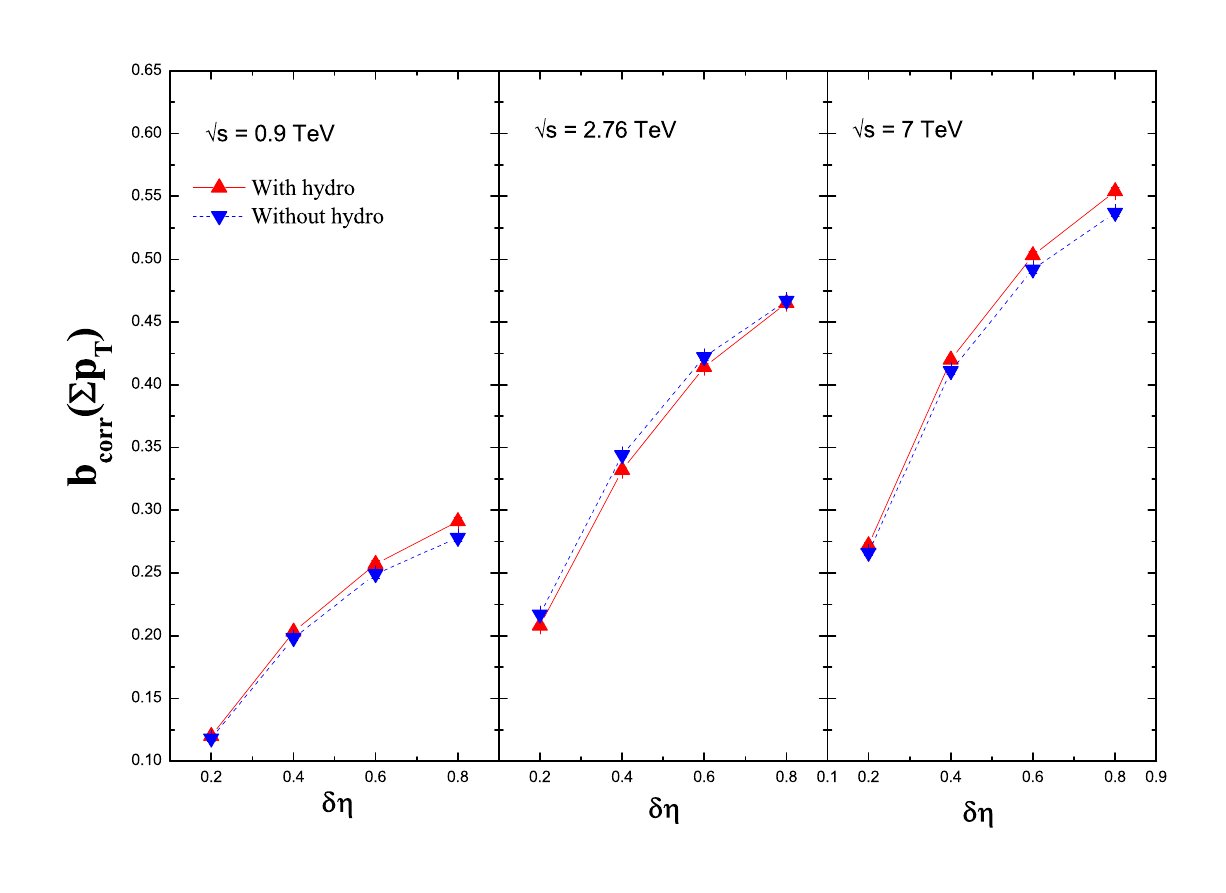}
\caption{(Color online) Dependence of $b_{corr} (\Sigma p_{T})$ on $\delta\eta$ for $\eta_{gap}$ = 0 in EPOS3 generated $pp$ 
events at $\sqrt{s}$ = 0.9, 2.76, and 7 TeV.}
\label{fig14}
\end{figure}

The nonlinear dependence of FB summed-$p_{\rm T}$ correlations on $\delta\eta$ is evident from Fig.~\ref{fig14} for EPOS3 
generated events. Similar to FB multiplicity correlation, we may think of the dominance of SRC component results in the nonlinear 
increase of  summed-$p_{\rm T}$ FB correlation.

\begin{figure}[hbt!]
\centering
\includegraphics[scale=0.45,keepaspectratio]{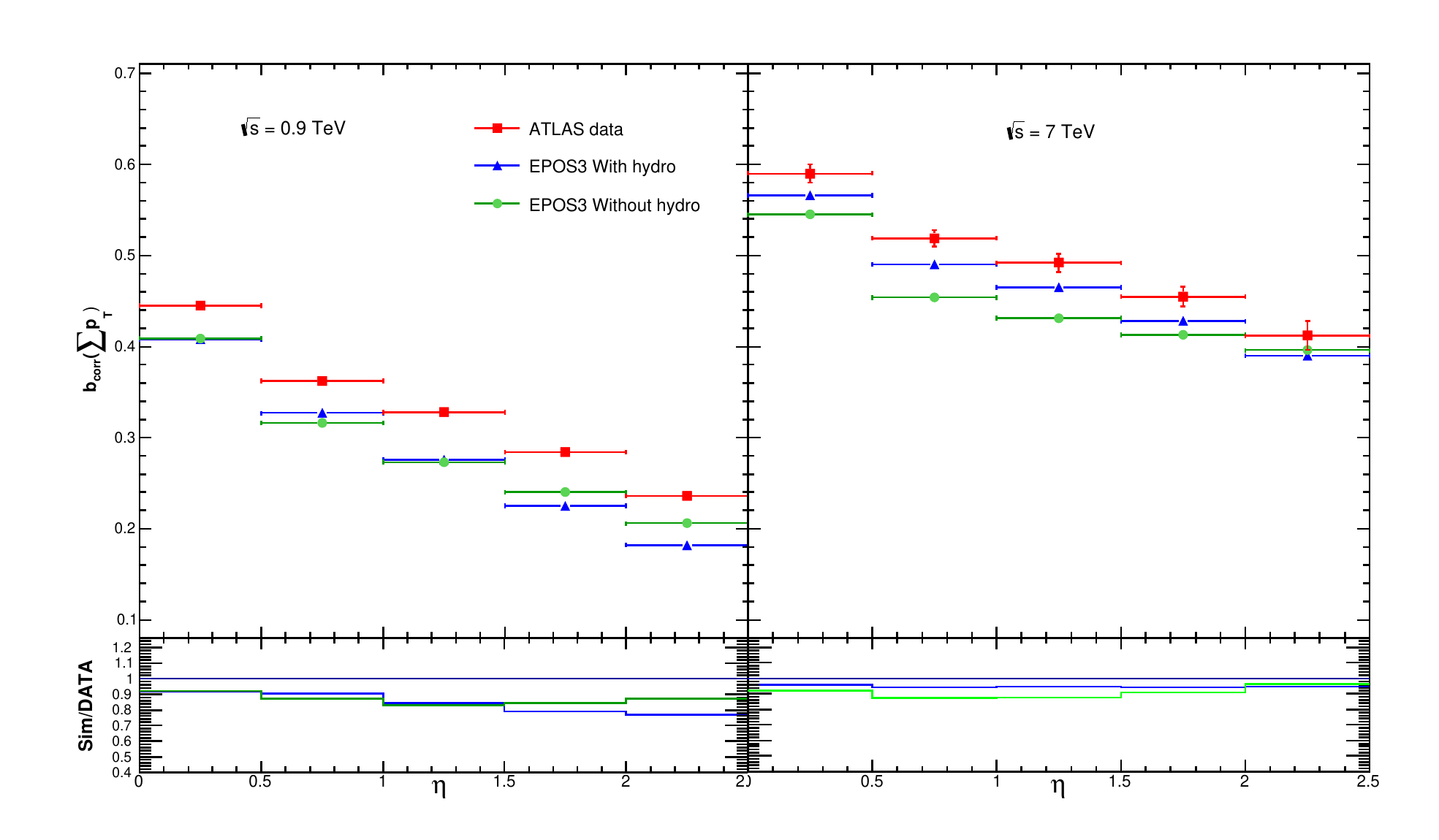}
\caption{(Color online) (Upper panel) Forward-backward summed-$p_{\rm T}$ correlation in symmetrically opposite $\eta$ intervals for 
EPOS3 simulated events with and without hydro at $\sqrt{s}$ = 0.9 and 7 TeV compared with ATLAS data. (Lower panel) 
Ratio of simulated events to ATLAS data.}
\label{fig15}
\end{figure}

Figure~\ref{fig15} shows that the $\eta_{gap}$ dependence of $b_{corr} (\Sigma p_{T})$ in the symmetrically 
opposite $\eta$ windows of equal width ($\delta\eta$ = 0.5) agrees with that of FB multiplicity correlation. For two different energies 0.9 
and 7 TeV,  we see that in comparison to 0.9 TeV, EPOS3 with hydro events is more comparable to data in 7 TeV. 

\begin{figure*}[hbt!]
\centering
\includegraphics[scale=0.32,keepaspectratio]{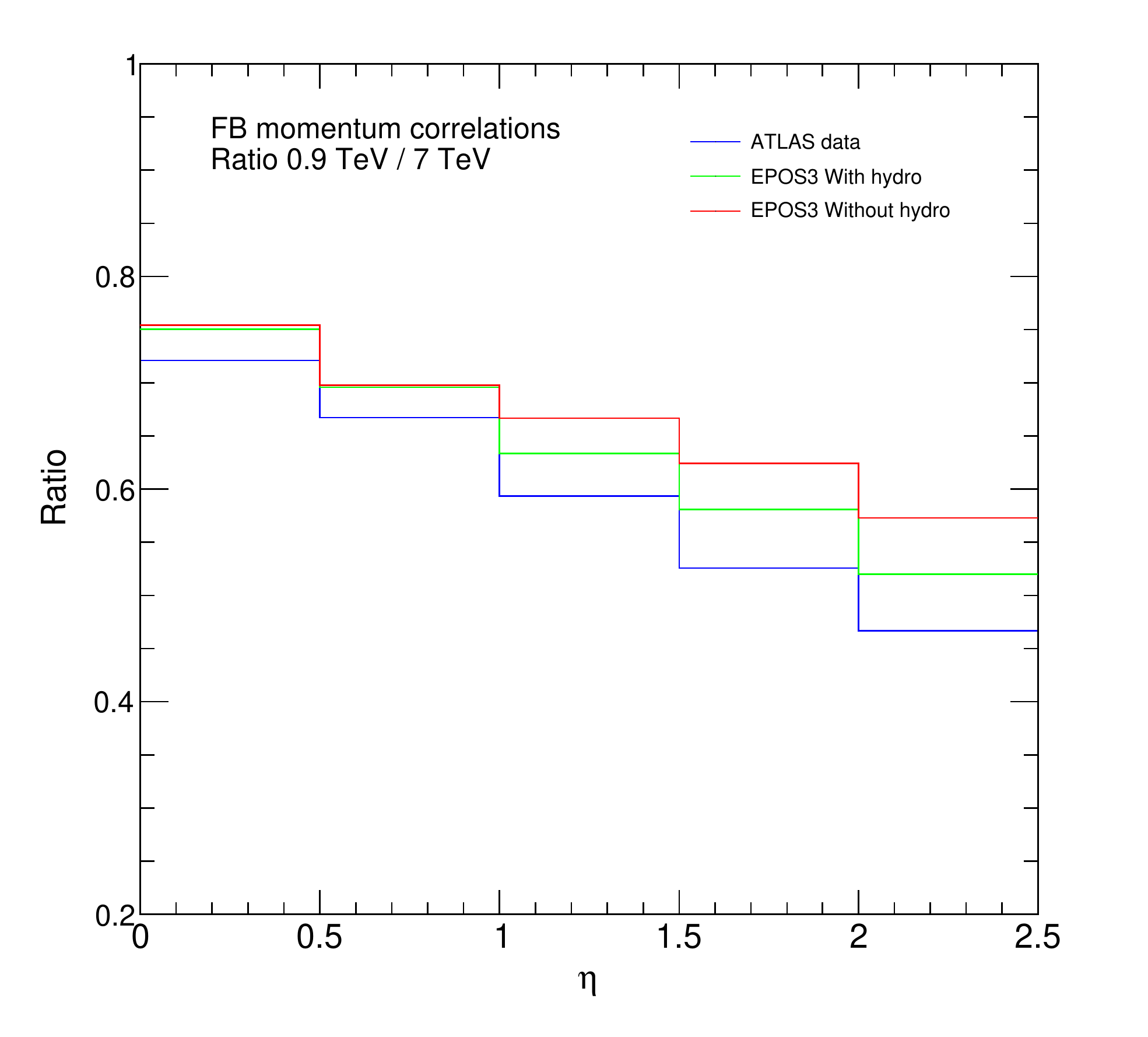}
\includegraphics[scale=0.58,keepaspectratio]{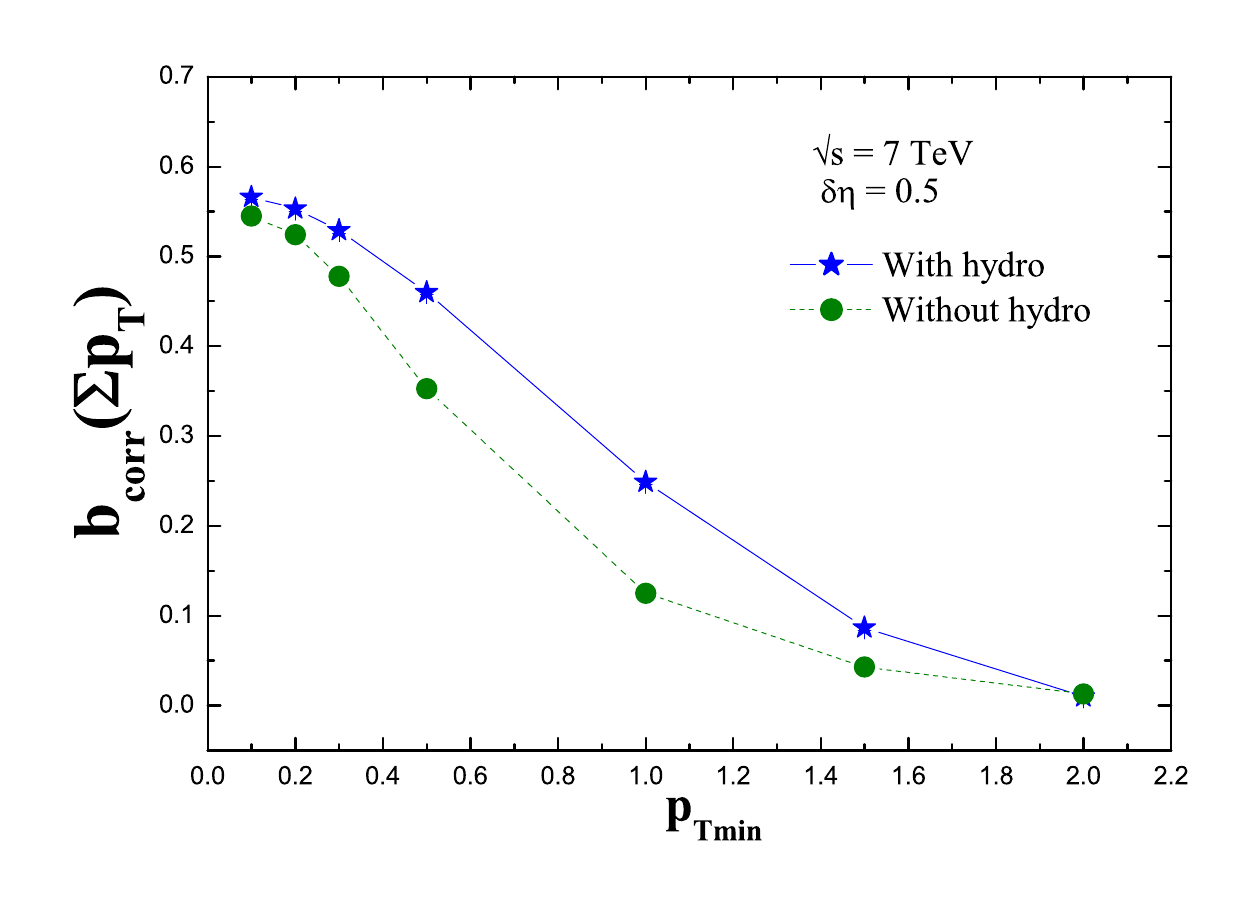}
\caption{(Color online) (Left) Ratio of forward-backward summed-$p_{\rm T}$ correlation for 0.9 TeV results to the 7 TeV results. 
(Right) Forward-backward summed-$p_{\rm T}$ correlation as a function of $p_{\rm T_{min}}$  for window width $\delta\eta$ = 0.5.}
\label{fig16}
\end{figure*}

Energy dependence is exhibited in the left panel of Fig.~\ref{fig16}, and it also supports the possible inference as predicted in case 
of multiplicity correlation. It is observed from the right panel of Fig.~\ref{fig16} that similar to $b_{corr}$ (mult), $b_{corr} (\Sigma p_{T} )$ 
also decreases rapidly with the transition from soft processes to hard processes, i.e, with $p_{\rm T_{min}}$ for EPOS3 events with and 
without hydro at $\sqrt{s}$ = 7 TeV.

\begin{figure}[hbt!]
\centering
\includegraphics[scale=0.34,keepaspectratio]{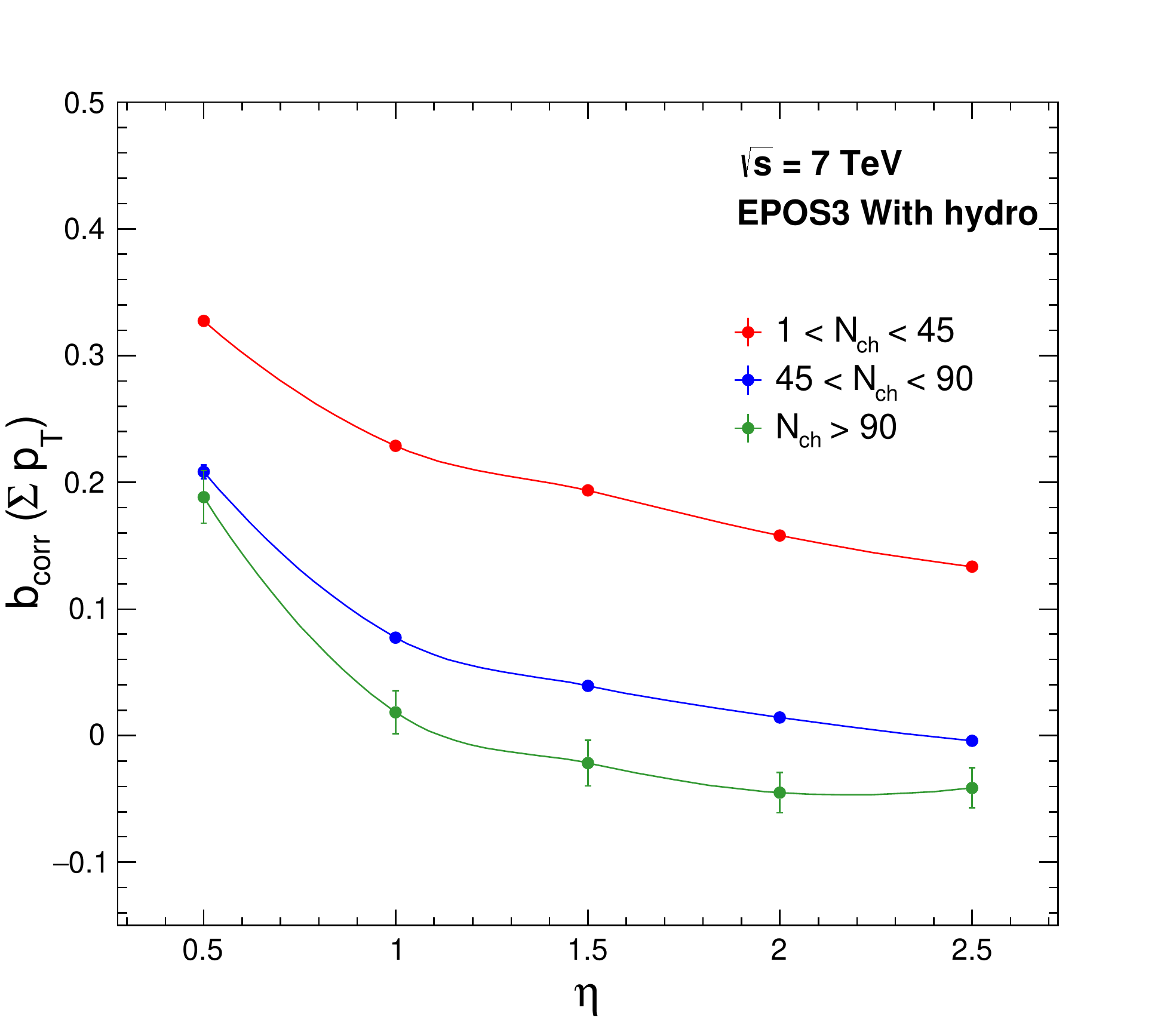}
\caption{Forward-backward summed-$p_{\rm T}$ correlations as a function of $\eta_{gap}$ for window width $\delta\eta$ = 0.5 
in different multiplicity range in EPOS3 simulated $pp$ events with hydro at $\sqrt{s}$ = 7 TeV.}
\label{fig17}
\end{figure}
So far we have used minimum-bias EPOS3 events with and without hydro to calculate the FB correlation strength. An attempt has been 
made to explore the multiplicity-dependent summed-$p_{\rm T}$ FB correlations in $pp$ collisions at 7 TeV using EPOS3 with hydro events. 
The reason behind choosing $b_{corr} (\Sigma p_{T} )$ over $b_{corr}$ (mult) can easily be understood from Sec.~\ref{sec2}. We divided 
the whole event sample into three nonoverlapping multiplicity regions: low (1 $ < N_{ch} <$ 45), mid (45 $ < N_{ch} <$ 90), and high 
($N_{ch} > $ 90), where $N_{ch}$ is the total number of charged particles, calculated following ATLAS kinematics (Sec.~\ref{sec4_2}). 
Figure~\ref{fig17} shows the FB summed-$p_{\rm T}$ correlations as a function of $\eta_{gap}$ for window width $\delta\eta$ = 
0.5 in those three multiplicity regions following the same ATLAS kinematics. We observe the similar decrease of correlation strength with increasing 
$\eta_{gap}$. Interestingly, we found that $b_{corr} (\Sigma p_{T} )$ decreases with increasing multiplicity at a fixed $\eta_{gap}$ and becomes 
lowest in high-multiplicity events. The decrease in correlation strength with increasing multiplicity could be due to the fact that, in EPOS3,
high-multiplicity events are generated via breaking of parent strings into a sequence of string segments producing a large string density, 
i.e, core. Such fusion of strings into core may lead to the smearing of correlation strength reflecting lower FB correlation 
in different $\eta$ window in high-multiplicity EPOS3 events. The negative values for $b_{corr} (\Sigma p_{T} )$ (anticorrelation) in 
high-multiplicity EPOS3 events in larger $\eta_{gap}$ could be due to lack of enough statistics. 
\section{Summary and Conclusions}\label{sec6}
We have seen that EPOS model successfully reproduces some basic features of particle production in $pp$ collisions at the LHC~\cite{ref40, ref41, ref42}. 
However, it fails to reveal few anomalous features in $pp$ collisions as well~\cite{ref32}. The present analysis highlights some important results 
and observations on long- and short-range correlations among produced charged particles 
in EPOS3 generated events at three center-of-mass energies $\sqrt{s}$ = 0.9, 2.76, and 7 TeV by exploring FB multiplicity and momentum correlation.\\

The study following ALICE kinematics reveals that
\begin{itemize}
\item {Both FB multiplicity and momentum correlation coefficients decrease slowly with the
increase of the gap between FB windows ($\eta_{gap}$) for each center-of-mass energy.}
\item{The $\eta_{gap}$ dependence of $b_{corr}$ maintains a nearly constant slop for all window widths in three center-of-mass energies.}
\item{The value of $b_{corr}$ increases nonlinearly with $\delta\eta$ for a fixed $\eta_{gap}$.}
\item{The pedestal value of $b_{corr}$ increases with collision energy.}
\end{itemize}
We observe that the general trends of $b_{corr}$ as a function of $\eta_{gap}$, $\delta\eta$, and collision energies as measured by ALICE Collaboration~\cite{ref17}, are fairly described by EPOS3 model.
Thus, our study corroborates ALICE experimental findings of FB correlations as well as predictions of different models, namely, Monte Carlo version of  QGSM~\cite{ref20}, Monte Carlo version of SFM~\cite{MCSFM1}, PYTHIA with different tunes~\cite{pythia6_default, pythi6tune, ref44}, and PHOJET~\cite{phojet} which qualitatively or quantitatively described the data.\\

The study following ATLAS kinematics reveals that
\begin{itemize}
\item{FB correlation varies strongly with $\eta_{gap}$ but weakly with the mean-$\eta$ value for a given pseudorapidity separation.}
\item{FB correlation decreases rapidly as minimum transverse momentum, $p_{\rm T_{min}}$ increases above a few hundred MeV.}
\item{FB summed-$p_{\rm T}$ correlation decreases as event multiplicity increases. A large deviation from minimum-bias study of $b_{corr} (\Sigma p_{T})$ with $\eta_{gap}$ is observed for high-multiplicity events.} 
\item{FB correlation strength increases with the increasing collision energy.}
\end{itemize}
It has been seen that the overall trend of above dependences is in agreement with the experimental results from ATLAS~\cite{ref4}. 
However, better agreement with ATLAS data has been noticed in the case of simulated $pp$ events with hydro for all FB window pairs except the most central one.

The observed rapid decrease of FB correlations with the increase of minimum transverse momentum, $p_{\rm T_{min}}$, as studied using EPOS3 simulated events, endorses the fact that at low $p_{T}$ values, partonic strings may uniformly fragment in the longitudinal direction but at higher $p_{T}$, particles may be associated with jets showing weak correlations between different jets~\cite{ref43}. Similar features are also predicted by the Monte Carlo version of String Fusion Model which anticipates that the decrease of correlation strength with the increase of $p_{\rm T_{min}}$ is related to the decrease of multiplicity restricting the overall string activities~\cite{MCSFM1}.

In addition to the minimum-bias study of EPOS3 simulated events, the multiplicity-dependent summed-$p_{\rm T}$ FB correlation shows significant changes in different multiplicity ranges. As discussed in~\cite{ref44}, the FB correlation strength can be sensitive to the changes of multiplicity and a significant variation in $b_{corr}$ has been reported in $pp$ collisions. The centrality dependence of FB correlations had already been predicted via different theoretical models including string fusion~\cite{SFM_cent}, string clustering framework~\cite{ColorCluster} for heavy-ion collisions. Such studies revealed that the long-range correlation strength increased from peripheral to central collisions. However, a strong suppression was observed in most central collisions which was explained in terms of suppression of color field fluctuations due to string fusion or interactions among cluster of color sources. Therefore, our observation on multiplicity-dependent $b_{corr} (\Sigma p_{T} )$ in $pp$ collisions adds more valuable information in this respect encouraging experimental measurements.

The energy dependence of FB correlation suggests that it might be due to the fact that the increase in long-range component of FB correlation is greater than its short-range component with the increase of multiple parton-parton interactions along with increasing center-of-mass energy~\cite{ref12}. 

It may be noted that the QCD-inspired multiparton interaction model like PYTHIA illustrated the FB multiplicity correlations by discriminating the power between different model tunes, particle production mechanisms, $p_{T}$ cuts and $\eta$ regions at $\sqrt{s}$ = 900 GeV~\cite{pythi6tune}. With further developed tunes, PYTHIA reproduces the trend of the FB correlation reasonably well as measured by ATLAS experiment~\cite{ref4}, though some of those tunes underestimate the FB correlation strength at high $\eta$ values. Furthermore, PYTHIA 8 tune A2 fails to describe the $N_{ch}$ dependency of SRC and LRC components as measured by ATLAS experiment in $pp$ collisions at $\sqrt{s}$ = 13 TeV~\cite{ref16}. Recent studies in PYTHIA with default color reconnection (CR) scheme~\cite{ref44} fail to explain the ALICE data, though somewhat better agreement is found with tuned CR scheme~\cite{ref45}. While PYTHIA with default CR scheme remains unsuccessful in explaining the LHC data in  terms of long-range correlations~\cite{ref46}, EPOS3 with hydrodynamical evolution of particles offers better agreement to the LHC data~\cite{ref30, ref33} in small systems ($pp/pA$). In view of recent correlation studies with EPOS3 model, our present study in FB multiplicity and summed-$p_{T}$ correlations using EPOS3 generated events will significantly contribute to the physics of multiparticle productions and interactions in high-energy $pp$ collisions.  

Overall, we may conclude that the hybrid Monte Carlo model, EPOS3 remains consistent in explaining the LHC data in terms of FB multiplicity and summed-$p_{\rm T}$ correlations qualitatively and explores the possible interplay between the soft and the hard processes in particle production in $pp$ collisions along with the variation of collision energy density. The study reflects that switching ON/OFF hydrodynamical evolution of bulk particles does not affect much the correlation strength rather multiparticle interactions and fluctuations plays important role in FB multiplicity correlations between particles in different $\eta$ windows as reported in various experiments and phenomenological models.

FB correlation strength can also be examined in different azimuthal windows in the $\eta-\varphi$ space selecting particles with different $p_{T}$ cuts.
This can be exploited as an effective tool for understanding and discriminating the source of the SRC and the LRC components~\cite{refFBtypes1}.
An exhaustive study in this light will be taken up separately in our future work. 
Furthermore, an extrapolation of such study would also be interesting in higher center-of-mass energy and multiplicity domain in $pp$ collisions  to test different aspects of the EPOS3 model.
\section*{Acknowledgements}
The authors express their gratitude to Dr. Subhasis Chattopadhyay, Variable Energy Cyclotron Centre (VECC) for providing academic support. 
The authors are thankful to the members of the grid computing team of VECC and cluster computing team of Department of Physics, Jadavpur University 
for providing uninterrupted facility for event generation and analyses. We also gratefully acknowledge the financial help from the DHESTBT, WB. One of 
the authors (J. M.) acknowledges DST-INDIA for providing fellowship under INSPIRE Scheme.  Another author (S. K.) acknowledges the financial 
support from UGC-INDIA Dr. D. S. Kothari Post Doctoral Fellowship under Grant No. F.4-2/2006(BSR)/PH/19-20/0039.

%
\end{document}